\documentclass[onecolumn,nofootinbib,nobibnotes,notitlepage]{revtex4-1}

\newcommand{\um}{\text{\textmu m}}		
\newcommand{\ie}{i.\,e.}				
\newcommand{\eg}{e.\,g.}				
\newcommand{\ea}{\textit{et al.}}		

\usepackage{amsmath}
\usepackage{xcolor}
\usepackage[english]{babel} 
\usepackage[utf8]{inputenc}
\usepackage{mathrsfs}
\usepackage{amsmath,amssymb,amsfonts}
\usepackage{graphicx}
\usepackage{hyperref}
\usepackage{array,multirow}
\usepackage{adjustbox}
\usepackage{cleveref}
\usepackage{dsfont}
\usepackage{xcolor}
\usepackage{textcomp}

\crefname{equation}{Eq.}{Eqs.}
\crefname{figure}{Fig.}{Figs.}
\crefname{section}{Sec.}{Secs.}
\crefname{table}{Tab.}{Tabs.}
\crefname{appendix}{Appx.}{Appx.}
\Crefname{equation}{Equation}{Equations}
\Crefname{figure}{Figure}{Figures}
\Crefname{section}{Section}{Sections}
\Crefname{table}{Table}{Tables}
\Crefname{appendix}{Appendix}{Appendices}

\begin{document}
\title{Coherent Fourier scatterometry reveals nerve fiber crossings in the brain}

\author{Miriam Menzel$^{1,*}$ and Silvania F.\ Pereira$^{2}$}

\affiliation{$^{1}$Institute of Neuroscience and Medicine (INM-1), Forschungszentrum Jülich, Wilhelm-Johnen-Straße, 52425 Jülich, Germany\\
	$^{2}$Optics Research Group, Department of Imaging Physics, Faculty of Applied Sciences, Delft University of Technology, Lorentzweg 1, 2628 CJ Delft, the Netherlands}

\email{m.menzel@fz-juelich.de}


\begin{abstract}
Previous simulation studies by Menzel \ea\ [Phys.\ Rev.\ X \textbf{10}, 021002 (2020)] have shown that scattering patterns of light transmitted through artificial nerve fiber constellations contain valuable information about the tissue substructure such as the individual fiber orientations in regions with crossing nerve fibers. Here, we present a method that measures these scattering patterns in monkey and human brain tissue using coherent Fourier scatterometry with normally incident light. By transmitting a non-focused laser beam ($\lambda = 633$\,nm) through unstained histological brain sections, we measure the scattering patterns for small tissue regions (with diameters of 0.1--1\,mm), and show that they are in accordance with the simulated scattering patterns. We reveal the individual fiber orientations for up to three crossing nerve fiber bundles, with crossing angles down to $25^{\circ}$. 
\end{abstract}
\maketitle


\section{Introduction}

With around 100 billion nerve cells on average \cite{herculano2009}, the brain is certainly the most complex organ in our body. Untangling this gigantic and highly complex nerve fiber network remains one of the biggest challenges in neuroscience. A precise knowledge about the nerve fiber pathways and connections is not only interesting for neuroanatomists; it is also a prerequisite for brain surgery and studies of neurological and mental disorders \cite{shi2017}.
Especially challenging is the correct reconstruction of densely packed, crossing nerve fibers. Due to an insufficient knowledge about nerve fiber crossings, tractography algorithms regularly misinterpret the course of nerve fiber pathways \cite{maierhein2017}. Even polarization microscopy --- one of the most powerful histological methods for mapping three-dimensional nerve fiber pathways in whole post-mortem brains with micrometer resolution \cite{MAxer2011_1,MAxer2011_2,reckfort2015} --- yields only a single fiber orientation for each measured tissue voxel and cannot reliably reconstruct fiber crossing points within a voxel \cite{dohmen2015}.

Recently, Menzel \ea\ (2020a) \cite{menzel2020} have shown that light scattering in brain tissue contains valuable information about the tissue substructure and can be used to reveal nerve fiber crossings. Using \textit{finite-difference time-domain (FDTD)} simulations and high-performance computing, they found that light transmitted through artificial nerve fiber constellations yields specific scattering patterns which contain structural information like the individual orientations of crossing nerve fiber bundles. The authors developed a dedicated simulation model for the imaging system and the inner structure of the nerve fibers, which allows for the first time to study complex brain tissue structures with FDTD simulations. The predictions of the simulations were successfully applied to identify regions with crossing nerve fibers in polarization microscopy measurements, and to develop a new imaging technique that measures the scattering of light under oblique illumination and has the potential to reconstruct the nerve fiber orientations for each image pixel of a brain section, also in regions with crossing fibers.

However, the scattering measurement can only be performed for a limited number of scattering angles, \ie\ only a small part of the full scattering pattern is considered, so that the angular accuracy of the determined fiber orientations is still limited ($\geq 15^{\circ}$ \cite{menzel2020}). To further develop this promising imaging technique, it is crucial to know to what extend the simulated scattering patterns are reliable.
Although the above findings suggest that the simulations make valid predictions, they can only be considered as an indirect validation of the simulation results. 
As discussed in \cite{menzel2020}, a direct validation of the simulation approach to correctly model brain tissue (scattering) properties is still missing.

In this paper, we develop a method that allows for the first time to measure the scattering patterns in brain tissue and provides a direct validation for the FDTD simulations of light scattering in brain tissue samples.
The method is based on \textit{coherent Fourier scatterometry} -- a proven method to study light scattering in non-biological, periodic samples \cite{kumar2014,gawhary2011}. Here, we modify the technique and apply it for the first time to measure light scattering in brain tissue.
We demonstrate both in a model system of crossing nerve fiber bundles (human optic chiasm) and in whole brain sections (vervet monkey), that the measured scattering patterns reveal the (in-plane) nerve fiber orientations with $< 1^{\circ}$ accuracy and resolve crossing angles down to $25^{\circ}$. The measured scattering patterns correspond very well to the simulated scattering patterns in \cite{menzel2020}, hence validating the employed simulation approach.


\section{Material and methods}

This section provides the background information and methods for this paper. We briefly describe how the simulated scattering patterns were generated (\cref{sec:simulation}), introduce our method for measuring scattering patterns in brain tissue with coherent Fourier scatterometry (\cref{sec:brain-samples,sec:measurement,sec:localization-laser}), and finally describe the evaluation of the scattering patterns (\cref{sec:evaluation}).


\subsection{Simulation of scattering patterns}
\label{sec:simulation}

The simulation studies by Menzel \ea\ (2020a) \cite{menzel2020} have shown that the distribution of light transmitted through artificial nerve fiber constellations reveals the substructure of the sample like the individual orientations of crossing fibers. \Cref{fig:scatterometry_vs_sim}(c) shows such a simulated scattering pattern for two crossing fiber bundles with $90^{\circ}$ crossing angle (adapted from \cite{menzel2020}, Fig.\ 7b).
Details about the simulation studies can be found in \cite{menzel2020,MMenzel}. Here, we provide only a brief summary of how the simulated scattering patterns were generated:
The propagation of light through the sample, \ie\ the electric and magnetic field components in space and time, were computed by a massively parallel 3D Maxwell solver (software TDME3D$^{\rm TM}$ \cite{wilts2014,michielsen2010}) based on a conditionally stable finite-difference time-domain (FDTD) algorithm \cite{taflove,menzel2016}. The algorithm discretizes space and time, and approximates Maxwell's equations by second-order central differences (see De Raedt \cite{deRaedt} for more details). The simulations were performed with a plane, coherent light wave ($\lambda = 550$\,nm) with circular polarization and normal incidence. The authors studied different configurations of myelinated nerve fibers (modeled by an inner axon surrounded by a myelin sheath with different refractive indices), within a volume of $30 \times 30 \times 30$\,\um$^3$ and with an average fiber diameter of $1$\,\um. After propagating through the sample, the intensity distribution of the scattered light was computed on a hemisphere behind the sample and projected onto the xy-plane, yielding a simulated scattering pattern as shown in \cref{fig:scatterometry_vs_sim}(c). 


\subsection{Measurement of scattering patterns with coherent Fourier scatterometry}
\label{sec:measurement}

To measure scattering patterns of brain tissue, we designed a measurement setup in the style of the simulations. 
In the following, we describe the basic principles of the measurement.
For more details about the setup and manufacturer information, the reader is referred to \cref{sec:scatterometry}. 

\begin{figure}[h!]
	\includegraphics[width=0.8\textwidth]{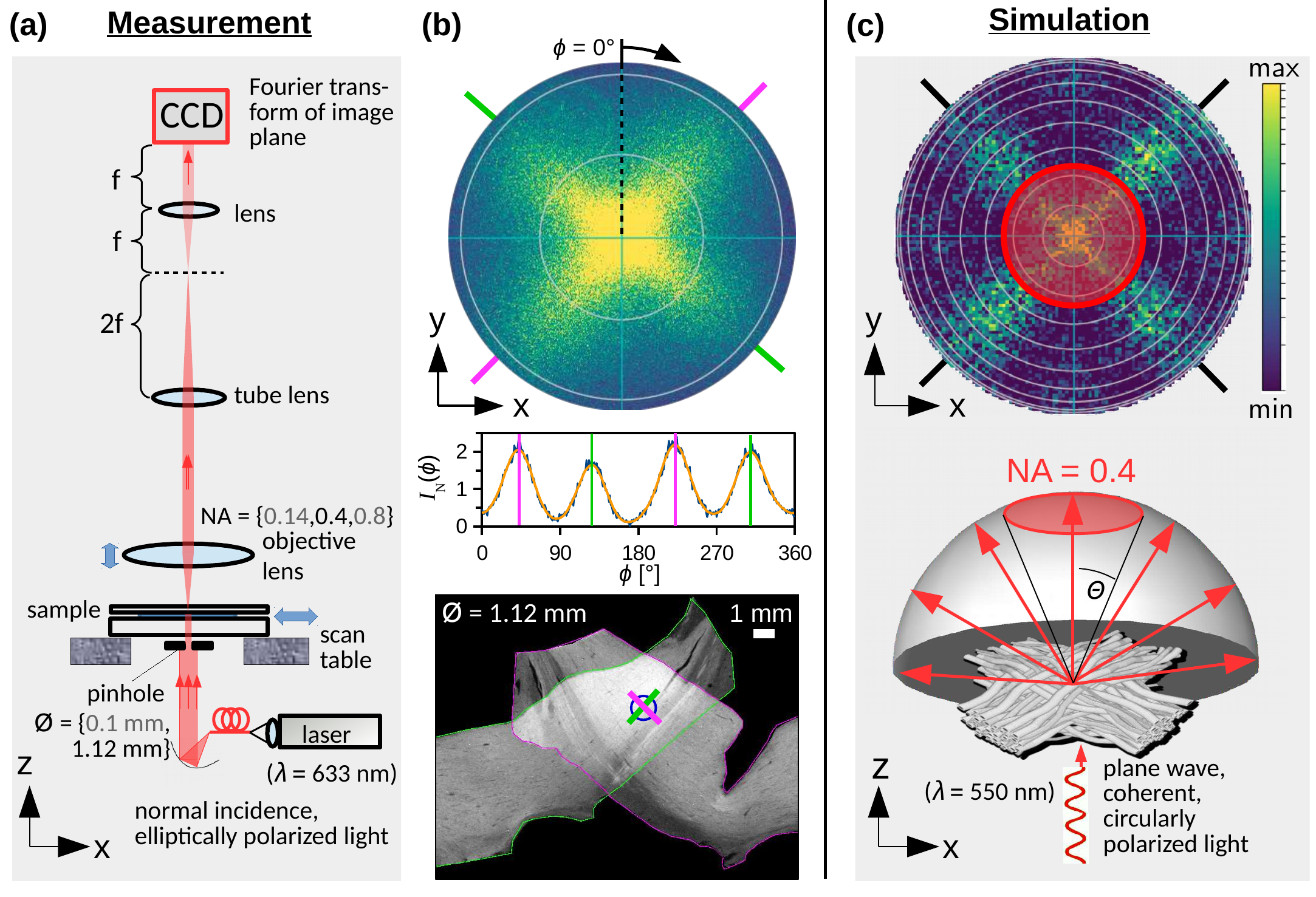}
	\caption{Scatterometry measurement vs.\ simulation: \textbf{(a)} Setup to measure scattering patterns of a brain section. Non-focused, normally incident laser light ($\lambda = 633$\,nm) is transmitted through the sample. The diameter of the laser beam is determined by a pinhole (with diameter $\varnothing = 0.1$\,mm or 1.12\,mm); the sample can be moved with micrometer screws in the x/y-direction. Different objective lenses with different numerical apertures (NA = \{0.14, 0.4, 0.8\}) are available. A camera (CCD) in the back-focal plane records the Fourier transform of the image plane (scattering pattern) for a given exposure time $t$. (The focal length of the lens in front of the camera is $f = 8$\,cm, the focal length of the tube lens is $2f = 16$\,cm.) \textbf{(b)} Scattering pattern and normalized polar integral obtained from a scatterometry measurement of a tissue region ($\varnothing = 1.12$\,mm, NA = 0.4, $t$ = 30\,ms) containing two crossing sections of human optic tracts. The magenta and green lines around the scattering pattern (top image) indicate the positions of the peaks. The dark-field microscopy image of the sample (bottom image) shows the measured tissue region (blue circle) and the predominant orientations of the nerve fibers (green/magenta lines), which are perpendicular to the determined peak positions. \textbf{(c)} Generation of simulated scattering pattern. A plane, coherent light wave ($\lambda = 550$\,nm) with circular polarization is transmitted through an artificial nerve fiber constellation (here: two crossing fiber bundles). The propagation of light is computed by an FDTD algorithm \cite{menzel2020}. The scattering pattern (top image) shows the distribution of scattered light intensity on a hemisphere behind the sample, projected onto the xy-plane. In the measurement, the maximum scattering angle $\varTheta$ that can be measured is limited by the numerical aperture of the objective lens (NA = $\sin\varTheta$) so that only the central area of the scattering pattern can be recorded (indicated by the red circle). The rings in the scattering pattern indicate steps of $\Delta\theta = 10^{\circ}$ (from $0^{\circ}$ in the center to $90^{\circ}$ for the outer ring); for NA = 0.4, only scattering angles up to $\varTheta = 23.6^{\circ}$ are collected. (The simulated scattering pattern was taken from \cite{menzel2020}, Fig.\ 7b, licensed under CC BY 4.0.)}
	\label{fig:scatterometry_vs_sim}
\end{figure}

The measurement setup is similar to the one by Kumar \ea\ (2014) \cite{kumar2014} to perform coherent Fourier scatterometry on printed gratings. While those authors focused the laser light onto the sample and measured the reflected light, we used a collimated beam to generate an approximately plane coherent light wave, and measured the transmitted light through the sample ($30$--$60$\,\um\ thin brain section), see \cref{fig:scatterometry_vs_sim}(a). The light was normally incident on the sample, and the scattered transmitted light was collected by a microscope objective. 
In contrast to the simulation, the light (generated by a helium-neon laser) has a wavelength of 633\,nm and is elliptically polarized (see \cref{sec:polarization}). However, it was shown that wavelength and polarization have no significant impact on the resulting scattering patterns (see \cite{menzel2020} and \cref{sec:polarization}).
To obtain the scattering pattern for a specific tissue region, the diameter of the laser beam was controlled by a circular pinhole (with diameter $\varnothing$) placed right below the sample. To avoid diffraction artifacts and ensure that the sample is illuminated by an approximately plane wave, the diameter of the pinhole should be much larger than the wavelength (the smallest pinhole diameter used in this study is 100\,\um). 
To measure different tissue regions, the sample was placed on a specimen stage that can be moved in the x/y-direction with micrometer screws.
The scattered light behind the sample was collected by an objective lens and measured by a CCD camera which was positioned in such a way that it records the Fourier transform of the image plane.
The resulting image is a scattering pattern (cf.\ \cref{fig:scatterometry_vs_sim}(b) on top) limited by the numerical aperture of the objective lens, as indicated in \cref{fig:scatterometry_vs_sim}(c) by the red circle.  

The measurements were performed for different beam diameters ($\varnothing = \{0.1$, 1.12\}\,mm), numerical apertures (NA = \{0.14, 0.4, 0.8\}), and exposure times ($t = \{10$, 30, 50, 150, 300, 600\}\,ms). The measurement parameters ($\varnothing$, NA, $t$) for all investigated brain tissue samples are listed in \cref{tab:table} in \Cref{sec:scatterometry-samples}.

\subsection{Localization of laser beam on the sample}
\label{sec:localization-laser}

\begin{figure}[htbp]
	\centering\includegraphics[width=0.65\textwidth]{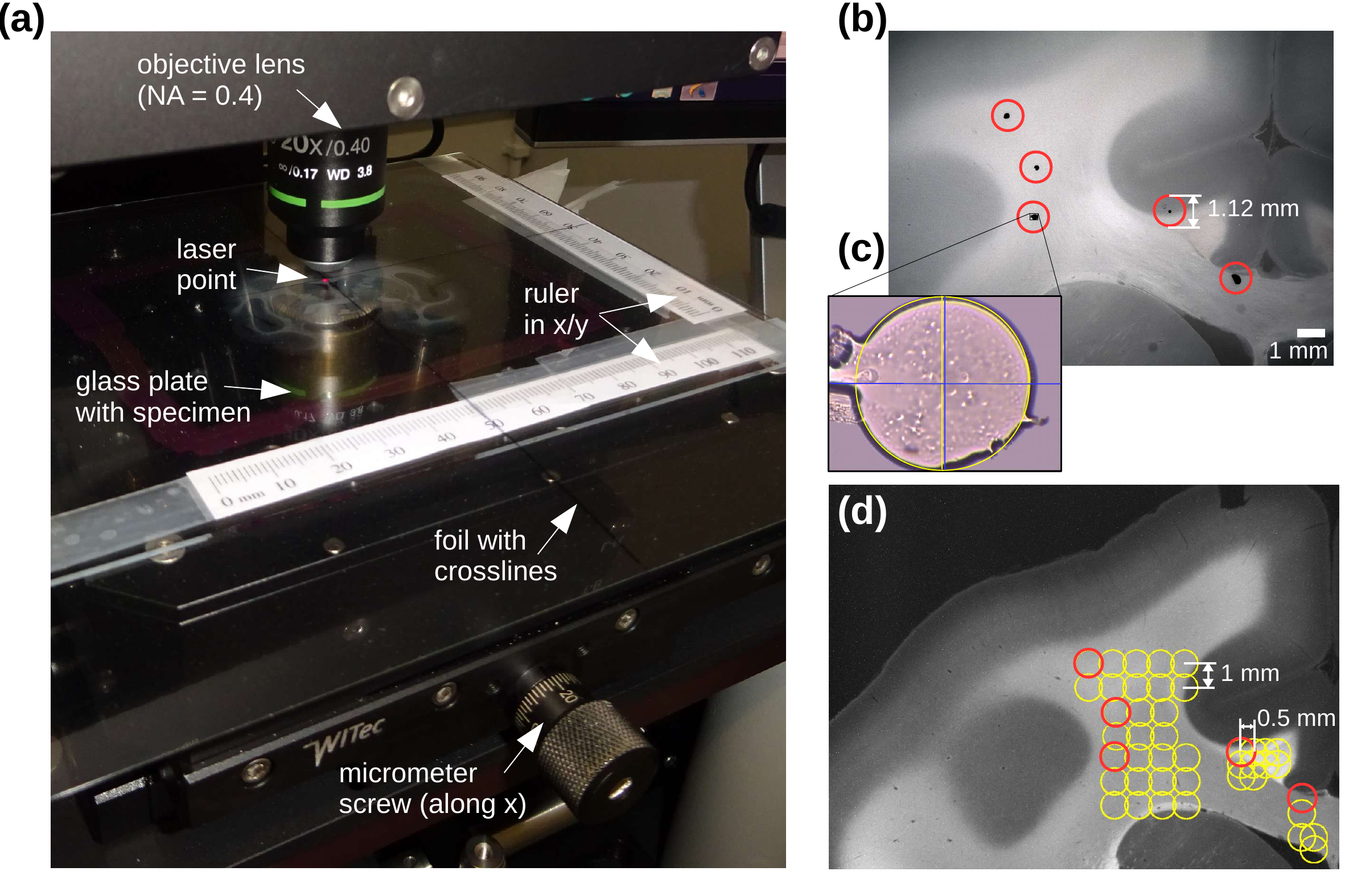}
	\caption{Localization of the laser point, shown exemplary for a coronal monkey brain section (vervet brain, section no.\ 458): \textbf{(a)} Photograph of the scan table during the scatterometry measurement ($\varnothing = 1.12$\,mm, NA = 0.4). The position of the laser beam on the brain section can be roughly determined by rulers and a transparent foil with cross-lines. To record scattering patterns of different tissue regions, micrometer screws were used to move the sample in steps of 0.5\,mm or 1\,mm in the x/y-directions. \textbf{(b)} Right before the scatterometry measurement, the starting points were marked on the cover glass with a pen (black dots) and the brain section was scanned with a digital microscope for reference (with aligned x/y-axes). The red circles indicate the laser beam (with 1.12\,mm diameter) used in the measurement. \textbf{(c)} To find the starting points, a bright-field image of the sample was recorded (with the setup shown in \cref{sec:scatterometry}, Fig.\ \ref{fig:scatterometry-setup}(a)) and the sample was moved until the image center (\ie\ the center of the laser beam, blue cross) lies in the center of the marked point (yellow cross). \textbf{(d)} The dark-field microscopy image of the same brain section was aligned with the image of the digital microscope, and the initial laser point positions (red circles) were transferred. According to how the sample was moved during the scatterometry measurement, the initial circles were translated in steps of 0.5\,mm or 1\,mm in the x/y-direction (yellow circles).}
	\label{fig:scat-point}
\end{figure}

In order to compare the measured scattering patterns to anatomical structures in the brain section, we need to determine the location of the laser beam on the sample during the scatterometry measurement. Due to the high magnification of the objective lens, the camera captures only a small portion of the sample (see Fig.\ \ref{fig:scat-point}(c) for NA = 0.4). Even when using rulers and a transparent foil with crosslines (see Fig.\ \ref{fig:scat-point}(a)), the position of the laser beam on the sample can only be roughly determined by eye. To accurately identify the measured tissue region, the starting points (initial positions of the laser beam) were marked on the cover glass of the sample with a pen (see black dots in Fig.\ \ref{fig:scat-point}(b)), and the sample was measured with a digital microscope (Keyence VHX 6000) ensuring that the borders of the glass plate are aligned with the x/y-axes of the microscope stage. In the scatterometry measurement, the sample was aligned with the scan table and moved until the laser beam was exactly located on one of the starting points\footnote{The scattering patterns are not influenced by the marked points (measurements with/without marker yield similar scattering patterns).} (a bright-field image was recorded to check for alignment, see Fig.\ \ref{fig:scat-point}(c)). Beginning at one starting point, the micrometer screws were used to move the sample in steps of 0.5\,mm or 1\,mm in the x/y-directions, and a scattering pattern was recorded for the different positions of the laser beam on the sample. 

As a measure of the overall scattering in the measured brain section, dark-field microscopy images with non-normally incident light were recorded for all investigated samples prior to the scatterometry measurements. 
To identify the location of the measured tissue region in the dark-field microscopy image (\cref{fig:scat-point}(d)), the image was aligned with the one of the digital microscope, and the initial laser point positions (red circles in Fig.\ \ref{fig:scat-point}(b)) were transferred to the dark-field microscopy image. According to how the sample was moved during the scatterometry measurement, the initial circles were translated in steps of 0.5\,mm or 1\,mm in the x/y-direction (yellow circles in Fig.\ \ref{fig:scat-point}(d)).


\subsection{Brain tissue samples}
\label{sec:brain-samples}

The brain tissue samples investigated in this study are two 60\,\um\ thin, coronal sections of a vervet monkey brain, and 30\,\um\ and 60\,\um\ thin sections of a human optic chiasm (cut along the fiber tracts of the visual pathway). 
To obtain well-defined samples with two or three crossing fiber bundles/layers, the sections of the optic chiasm were cut into two parts at the median line (left and right), and the sections of the optic tracts were manually placed on top of each other with different crossing angles (see \cref{fig:scatterometry_samples}(b) in \cref{sec:scatterometry-samples}). The sections of the optic tract are particularly well suited as model systems because they contain many parallel (myelinated) nerve fibers with well-defined orientations. All brain samples were placed on a glass plate, embedded in glycerin solution, and covered by a cover slip. The preparation of the brain samples is described in \cref{sec:preparation} in more detail.

\Cref{fig:scatterometry_samples}(c) shows the dark-field microscopy images of all investigated brain sections (vervet brain and human optic tracts) and the tissue regions (colored circles) that were measured with scatterometry. As described in \cref{sec:localization-laser}, the position of each tissue region is uniquely identified by x/y-coordinates (the origin is the starting point of the measurement, marked by $\ast$). For example, the upper right green circle in \cref{fig:scatterometry_samples}(c) is referred to as: \textit{``Vervet Brain (section 493), cr1, x = 3\,mm, y = -1\,mm''}.
The dates of the dark-field microscopy and scatterometry measurements can be found in \cref{tab:table} in \cref{sec:scatterometry-samples}.


\subsection{Evaluation of scattering patterns}
\label{sec:evaluation}

The top image in \cref{fig:scatterometry_vs_sim}(b) shows the measured scattering pattern for a tissue region containing two (almost perpendicularly) crossing sections of optic tracts. The blue circle in the bottom image indicates the position of the laser beam during the measurement, the magenta and green lines indicate the predominant orientations of the nerve fibers in the measured tissue region. The magenta and green lines around the scattering pattern indicate which scattering peak was caused by which fiber bundle.

In order to quantitatively evaluate the measured scattering patterns and determine the nerve fiber orientations from the scattering peaks, we computed the \textit{azimuthal integral} $I(\theta)$ and the \textit{polar integral} $I(\phi)$ of the scattering patterns (see \cref{sec:evaluation-scatteringpattern} for more details). The graph in \cref{fig:scatterometry_vs_sim}(b) shows the polar integral of the measured scattering pattern: The intensity values were integrated from the center to the outer border of the pattern (see black dashed line) for a certain azimuthal angle $\phi$ --- taking the geometry of the projected hemisphere into account --- and plotted against $\phi$ (starting on top and rotating clock-wise, see black arrow). 
To determine the position of the peaks, the polar integrals were \textit{smoothed} out, using a Savitzky-Golay filter \cite{savitsky1964} with 45 sampling points and a second order polynomial (see \cref{sec:evaluation-scatteringpattern}). The graph shows both the original curve (blue) and the smoothed curve (orange), together with the determined peak positions (vertical colored lines).

For better comparison, the polar integrals were \textit{normalized}: The background noise (minimum detected intensity value) was subtracted from the polar integrals, and the resulting intensity values were divided by the average of the signal: 
\begin{align}
I_{\rm N}(\phi) = \frac{I(\phi) - I(\phi)_{\rm min}}{\overline{I(\phi)}}.
\end{align}

In Dataset 1 (Ref.\ \cite{dataset}), the reader can find all measured scattering patterns and corresponding azimuthal/polar integrals ($I(\theta)$ and $I_{\rm N}(\phi)$) for the tissue regions shown in \cref{fig:scatterometry_samples} (labeled by brain section, brain region, and x/y-coordinates --- as described in \cref{sec:brain-samples}).


\section{Choice of system parameters}

In order to determine the correct in-plane orientations of crossing nerve fibers, the peaks in the smoothed polar integrals should be determined as precisely as possible.
An important quality measure in this context is the \textit{noise} $N$. The smaller the noise, the more reliable are the measured signal and the determined peak positions. The noise is defined by the difference between the original polar integral $I(\phi)$ and the smoothed polar integral $\tilde{I}(\phi)$, divided by the amplitude of the smoothed polar integral:
\begin{align}
N = \frac{I(\phi)-\tilde{I}(\phi)}{\tilde{I}(\phi)_{\rm max} - \tilde{I}(\phi)_{\rm min}}.
\label{eq:noise}
\end{align}

The \textit{signal-to-noise ratio} $S/N$ is defined as the standard deviation $\sigma\{x\}$ of the signal (amplitude of the smoothed polar integral) divided by the noise: 
\begin{align}
S/N = \sigma\left\{\frac{\tilde{I}(\phi)_{\rm max} - \tilde{I}(\phi)_{\rm min}}{I(\phi)-\tilde{I}(\phi)}\right\}.
\label{eq:SNR}
\end{align}


\subsection{Noise analysis}

\begin{figure}[t]
	\begin{center}
		\includegraphics[width=0.6\textwidth]{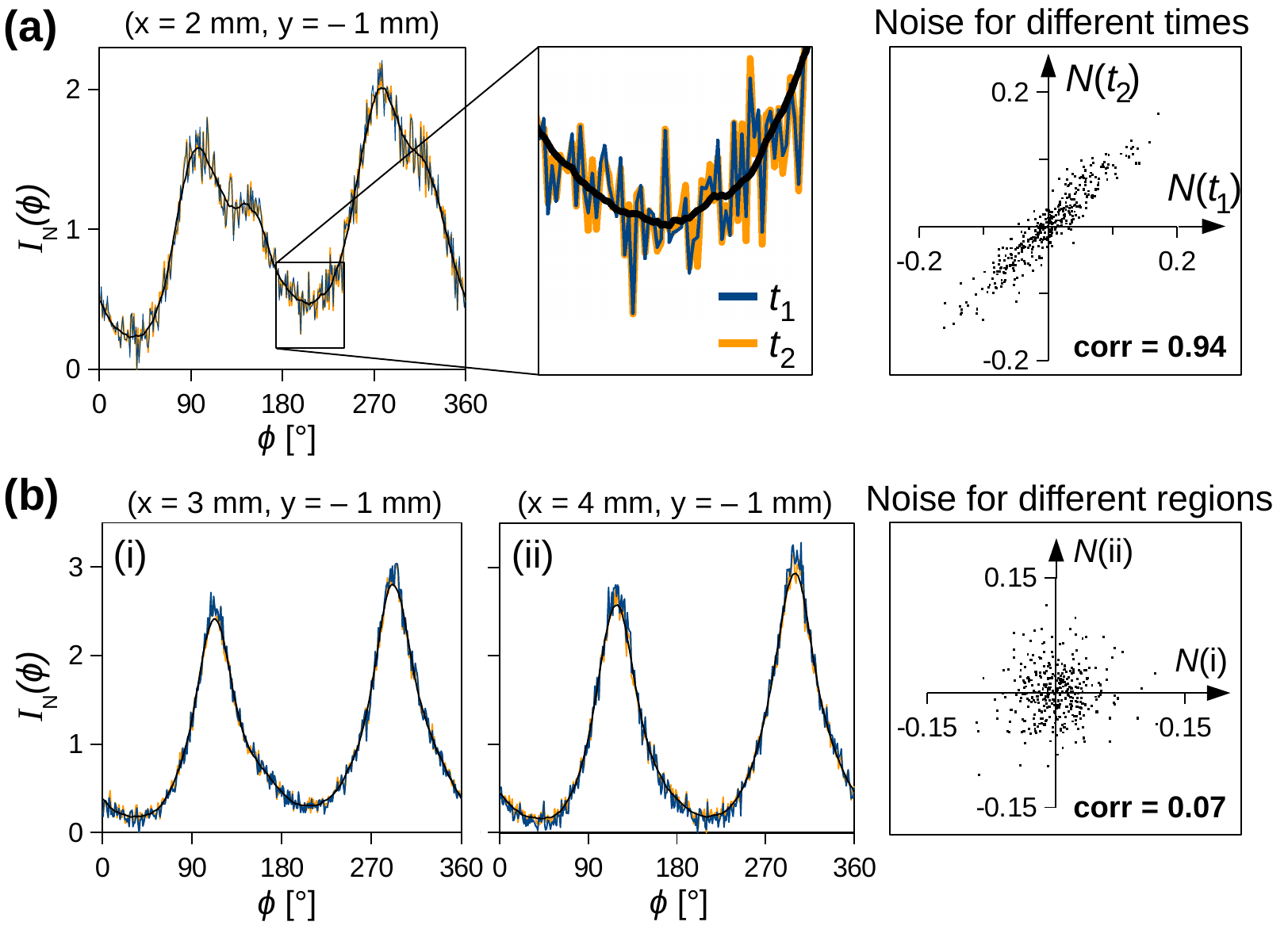}
		\caption{Noise measured for different tissue regions ($\varnothing = 1.12$\,mm) in the corona radiata of a coronal vervet brain section (section no.\ 458, cr1, x = \{2,3,4\}\,mm, y = -1\,mm; see \cref{fig:scatterometry_samples}(c)): \textbf{(a)} Polar integrals of the same tissue region measured at two different times $t_1$ and $t_2$ (blue/orange curves). The black curve shows the smoothed polar integral. The scatter plot on the right shows the noise (\cref{eq:noise}) for $t_2$ plotted against $t_1$. \textbf{(b)} Polar integrals for two similar, neighboring tissue regions ((i) and (ii)). The scatter plot shows the noise for (ii) plotted against (i). corr = correlation coefficient: ${\rm cov}(x,y)/(\sigma_x \sigma_y)$.}	
		\label{fig:noise}
	\end{center}
\end{figure}

To study the noise, five tissue regions in a vervet brain section (red shaded circles in \cref{fig:scatterometry_samples}(c)) were measured at two different times $t_1$ and $t_2$, with $(t_2 - t_1) > 10$\,min. The measurements were performed with the same beam diameter, numerical aperture, and exposure time ($\varnothing = 1.12$\,mm, NA = 0.4, $t = 30$\,ms).
For three of these tissue regions, \cref{fig:noise} shows the resulting polar integrals for $t_1$ and $t_2$ together in one plot (blue/orange curves). The zoomed-in area in \cref{fig:noise}(a) shows that the two curves --- although obtained from measurements at two different times --- correspond very well to each other. Not only are the smoothed curves (in black) almost identical, also the original zigzag curves before smoothing (blue/orange) are very similar to each other. The scatter plot on the right shows the noise (difference between original and smoothed curve for $\phi = \{0^{\circ}, 1^{\circ}, \dots, 359^{\circ}\}$) for the polar integral obtained at time $t_2$ plotted against the noise at time $t_1$. The correlation coefficient is very high (0.94).

\Cref{fig:noise}(b) shows the polar integrals for two similar, neighboring tissue regions. Although the smoothed polar integrals ((i) and (ii)) are very similar to each other, the noise is not correlated. The scatter plot on the right shows the noise of the blue curves plotted against each other; the correlation coefficient is very small (0.07).

This suggests that the fine structures (zigzag lines) in the original polar integrals (blue/orange curves) are caused by details in the underlying fiber structure, \eg\ the exact orientation/diameter of individual nerve fibers in a bundle, and not by time-dependent background noise or systematic noise in the measurement. Hence, small changes in the sample position, \ie\ in the position of the measured tissue region, might significantly change the detailed structure of the resulting scattering pattern (zigzag lines in the original polar integral), but not the overall structure (smoothed polar integral) which is related to the overall nerve fiber organization in the measured region. It can be noted that more inhomogeneous tissue regions with several different fiber orientations (like in \cref{fig:noise}(a)) have a higher level of noise and are more sensitive to small changes in the sample position than more homogeneous tissue regions with one dominant fiber orientation (like in \cref{fig:noise}(b)). Similar observations were also made in the simulations (see \cite{menzel2020}), where small changes in the simulation parameters or fiber configurations also caused small changes in the simulated scattering patterns, but the overall structures like the positions of the overall scattering peaks remain the same.
As we are here only interested in the overall fiber structure like the orientations of crossing fibers, we only considered the smoothed polar integrals of the measured scattering patterns.

\subsection{Choice of pinhole size and numerical aperture}
\label{sec:pinhole-size}

\begin{figure}[htbp]
	\includegraphics[width=0.9\textwidth]{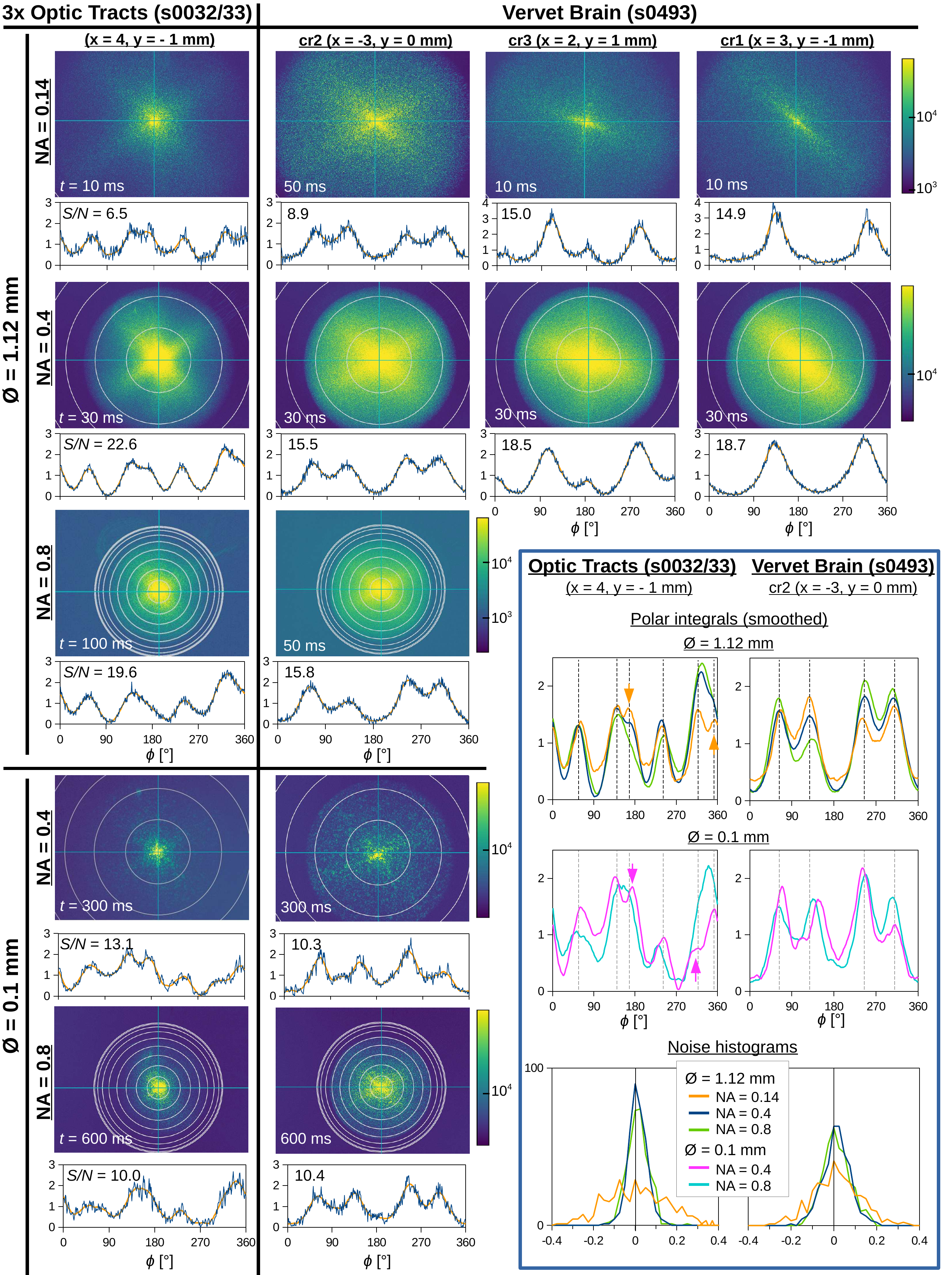}
	\caption{Scattering patterns of four different tissue regions in two brain tissue samples (three crossing sections of optic tracts and a coronal vervet brain section, cf.\ \cref{fig:scatterometry_samples}(c)) measured with different laser beam diameters ($\varnothing$), numerical apertures (NA), and exposure times ($t$). The rings in the scattering patterns indicate steps of $\Delta\theta= 10^{\circ}$ on the hemisphere. The graphs underneath the scattering patterns show the corresponding (smoothed) polar integrals and the signal-to-noise ratio ($S/N$) computed with \cref{eq:SNR}. The graphs in the blue box on the lower right show --- for two of the tissue regions --- the smoothed polar integrals for different numerical apertures and pinhole diameters in one plot (upper two rows), and the histograms of the noise computed with \cref{eq:noise} (lower row).}	
	\label{fig:pinhole-NA}
\end{figure}

To determine the optimum system parameters for the scatterometry measurement, several tissue regions (highlighted in \cref{fig:scatterometry_samples}(c) in different colors) were measured with different laser beam diameters ($\varnothing$ = \{0.1, 1.12\}\,mm), numerical apertures (NA = \{0.14, 0.4, 0.8\}), and exposure times ($t$ = 10--600\,ms).
\Cref{fig:pinhole-NA} shows the scattering patterns, (smoothed) polar integrals, and the computed signal-to-noise ratio ($S/N$, see \cref{eq:SNR}) for four different tissue regions: one region containing three crossing sections of optic tracts (human chiasm, section no. 32/33) and three regions containing (non-)crossing fibers in the corona radiata of a coronal vervet brain section (no.\ 493). The exact positions of the measured tissue regions are indicated by x/y-coordinates (cf.\ \cref{fig:scatterometry_samples}(c)).

The measured scattering patterns are limited by the numerical aperture of the objective lens: For NA = \{0.14, 0.4, 0.8\}, the maximum scattering angles are: $\varTheta = \{8.0^{\circ}, 23.6^{\circ}, 53.1^{\circ}\}$, respectively. The angular resolution, \ie\ the resolution in $k$-space, is higher for small numerical apertures. At the same time, the polar integrals are computed over a smaller distance in $\theta$ because the scattering patterns show only a small, inner part of the full scattering pattern.

The sections of the human optic tracts were measured more than 125 days after tissue embedding and the overall scattering of the tissue is less than for the vervet brain section, which was measured about 45 days after tissue embedding (see \cref{fig:scatterometry_samples}(c) and \cref{tab:table}). 
For weakly scattering tissue (optic tracts), the signal-to-noise ratio for NA = 0.14 is much less than for NA = 0.4 or 0.8, and the distribution of noise in the histogram is much broader. The highest signal-to-noise ratio ($S/N = 22.6$) was achieved for a beam diameter of 1.12\,mm, a numerical aperture of 0.4, and an exposure time of 30\,ms.
For strongly scattering tissue (vervet), the signal-to-noise ratio for NA = 0.8 is slightly larger than for NA = 0.4. When using small exposure times of 10\,ms and 30\,ms, the signal-to-noise ratios for NA = 0.14 and 0.4 are similar (around 15 and 18.5). Although $S/N$ is slightly lower, the peak positions can be more precisely determined for NA = 0.14 because the peak widths become smaller (see scattering patterns in the upper right of \cref{fig:pinhole-NA} in comparison to the scattering patterns in the row below).

When using a small beam diameter ($\varnothing = 0.1$\,mm) and a long exposure time (to get enough signal), the signal-to-noise ratio becomes smaller and the peaks are not as clearly defined as for a beam diameter of 1.12\,mm.
However, as a comparison of the smoothed polar integrals shows (see upper two rows in the blue box of \cref{fig:pinhole-NA}), the positions of the peaks for $\varnothing = 0.1$\,mm are still similar to the peak positions determined for $\varnothing = 1.12\,$mm (vertical dashed lines). This shows that the fiber orientations can also be determined for small tissue regions with diameters down to 100\,\um\ (\ie\ with a comparable order of magnitude as in the simulations).
For weakly scattering tissue (optic tracts), the peaks for NA = 0.4 are better visible (magenta arrows) than for NA = 0.8. The minimum peak distance is about $25^{\circ}$, showing that the scatterometry measurement can reveal the fiber orientations of crossing nerve fiber bundles with crossing angles down to $25^{\circ}$.

In summary: 
A small beam diameter ($\varnothing = 0.1$\,mm) allows to resolve more details in neighboring fiber structures, but also leads to a lower signal-to-noise ratio.
A small numerical aperture (NA = 0.14) and a short exposure time allow to resolve more details in the center of the scattering patterns and to distinguish closely neighboring scattering peaks (see orange arrows in \cref{fig:pinhole-NA}), but the signal-to-noise ratio is very low for weakly scattering brain tissue (optic tracts).
A high numerical aperture (NA = 0.8) and a long exposure time, on the other hand, allow to obtain more information from the borders of the scattering pattern, yielding more reliable peak positions and a slightly larger signal-to-noise ratio for strongly scattering tissue (vervet). For weakly scattering tissue (optic tracts), however, the signal-to-noise ratio for NA = 0.8 is still lower than for NA = 0.4.
As a compromise, we used a laser beam diameter of 1.12\,mm, a numerical aperture of 0.4, and an exposure time of $30$\,ms for all following studies.


\section{Results}


\subsection{Model system of crossing optic tracts}

\begin{figure}[t]
\includegraphics[width=0.8\textwidth]{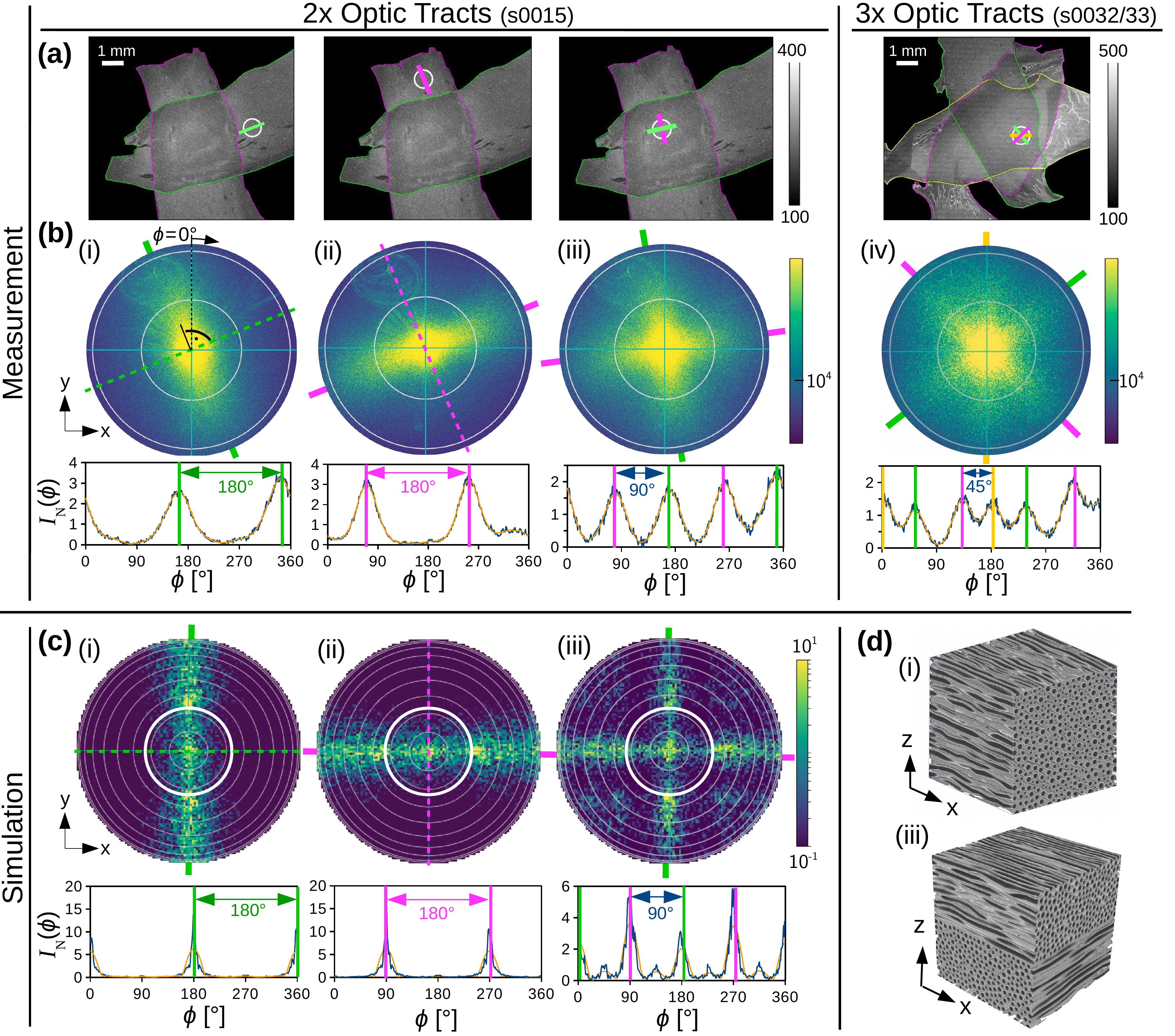}
	\caption{Measured vs.\ simulated scattering patterns for different crossing fiber layers: \textbf{(a)} Dark-field microscopy images of two and three crossing sections of optic tracts. The white circles show the tissue regions measured with scatterometry ($\varnothing = 1.12$\,mm, NA = 0.4,  $t =$ 30 ms), consisting of one ((i),(ii)), two (iii), and three (iv) crossing fiber layers. The outline of the optic tract sections is shown in different colors for better reference. The straight colored lines indicate the fiber orientations of the respective layers in the measured tissue region. \textbf{(b)} Measured scattering patterns and normalized (smoothed) polar integrals of the four tissue regions (i)--(iv) indicated in (a). The non-dashed, colored lines indicate the positions of the scattering peaks, the dashed colored lines (in (i),(ii)) the predominant orientation of the nerve fibers in the measured tissue region. \textbf{(c)} Simulated scattering patterns for parallel fibers oriented in the x-direction (i) and y-direction (ii), and two crossing fiber layers with $90^{\circ}$ crossing angle (iii). The graphs below show the normalized (smoothed) polar integral. \textbf{(d)} Artificial fiber constellations ($30 \times 30 \times 30$\,\textmu m$^3$) used to compute the simulated scattering patterns in (c). The scattering patterns in (c) and the fiber configurations in (d) were adapted from Menzel \ea\ (2020b) \cite{menzel2020-OSA}, Fig.\ 1(c), licensed under CC BY 4.0.}
	\label{fig:scatterometry_chiasm}
\end{figure}

\Cref{fig:scatterometry_chiasm} shows the measured scattering patterns and polar integrals of two and three crossing sections of optic tracts (human chiasm, sections no.\ 15 and 32/33, cf.\ \cref{fig:scatterometry_samples}) in comparison to the simulated scattering patterns of constellations with parallel fibers and two $90^{\circ}$-crossing fiber layers. 
As predicted by the simulations, the light is scattered perpendicularly to the predominant nerve fiber orientation.
In regions containing a single layer of fibers, \ie\ one section of optic tracts ((i) and (ii)), the scattering peaks are perpendicular to the predominant orientation of the nerve fibers in the layer (green/magenta dashed lines in the scattering patterns); the polar integrals show two distinct peaks that lie $180^{\circ}$ apart. 
In regions containing several crossing layers of fibers ((iii) and (iv)), each fiber layer generates two $180^{\circ}$-peaks (vertical colored lines in the polar integrals), which are perpendicular to the respective nerve fiber orientation (see colored lines in (a)). In a tissue region containing two $90^{\circ}$-crossing fiber layers (iii), this results in four distinct peaks that lie $90^{\circ}$ apart. In a tissue region with three $45^{\circ}$-crossing fiber layers (iv), this results in six distinct peaks that lie $45^{\circ}$ apart. The scattering pattern of two crossing fiber layers (iii) can be considered as a superposition of the scattering patterns (i) and (ii) of the individual, non-crossing fiber layers.
The scattering patterns were measured with NA = 0.4 and can only be compared to the inner part of the simulated scattering patterns\footnote{The simulated scattering patterns were only evaluated for NA = 1 because the resolution in $k$-space (\ie\ the number of scattering angles) is limited by computing time. Integrating over a smaller number of scattering angles (lower numerical aperture) would lead to a poor signal-to-noise ratio.} (marked by a white circle). Taking this into account, the measured and simulated scattering patterns look very similar. 
The minor peaks in the simulated scattering pattern of the two crossing fiber layers, which are also visible in the polar integral (iii), occur for larger scattering angles and are not expected to occur for NA = 0.4.  
\enlargethispage{1cm}


\subsection{Fiber architectures in whole brain section}

\begin{figure}[h!]
\includegraphics[width=0.8\textwidth]{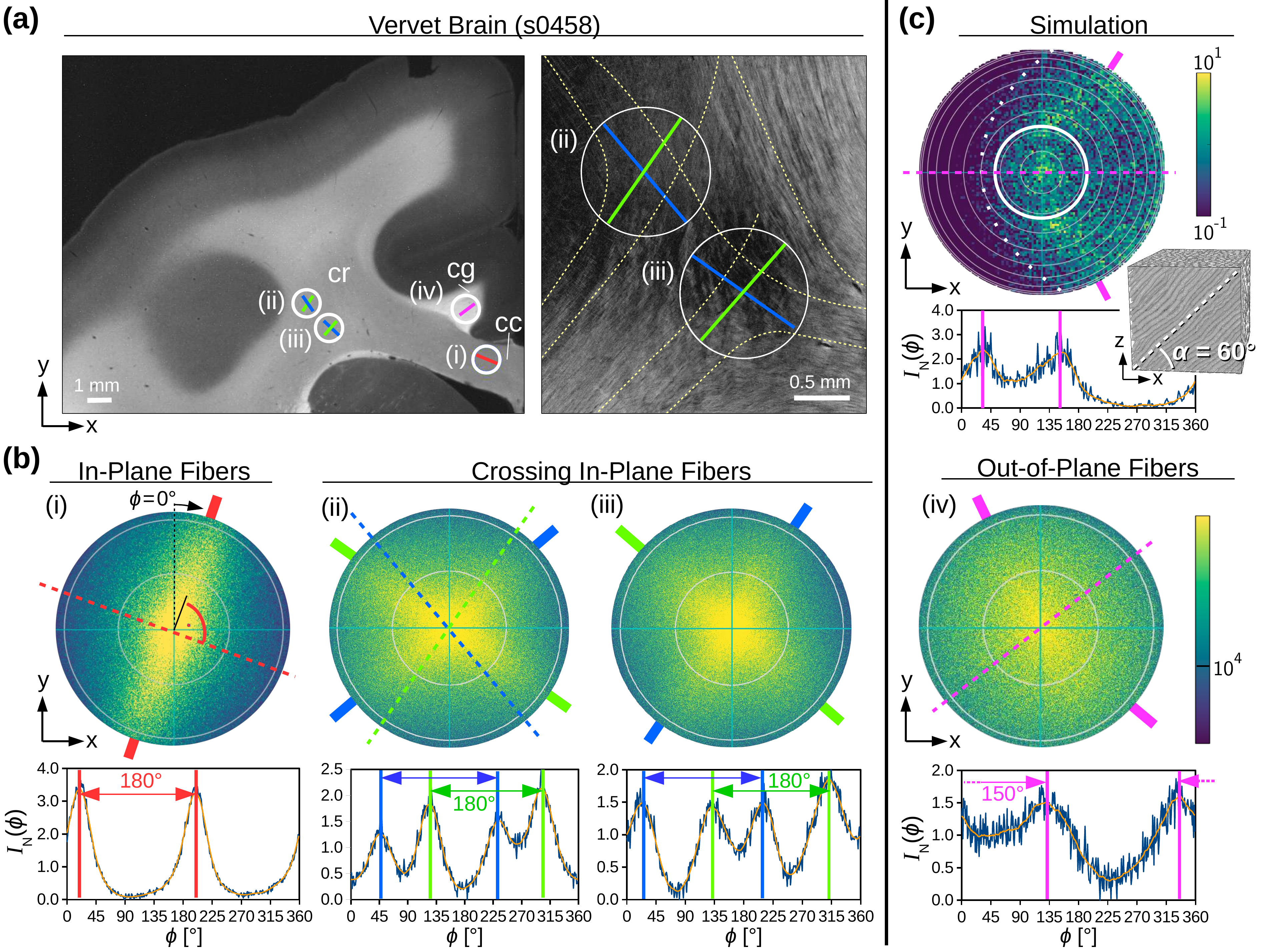}
	\caption{In-plane, crossing, and out-of-plane nerve fibers of a coronal vervet brain section studied with scatterometry: \textbf{(a)} The image on the left was obtained from a dark-field microscopy measurement of the left upper corner of the brain section (the whole brain section is shown in \cref{fig:scatterometry_samples}(a) in the Appendix). The white circles show the tissue regions measured with scatterometry ($\varnothing = 1.12$\,mm, NA = 0.4,  $t =$ 30 ms) with in-plane parallel (i), crossing ((ii),(iii)), and out-of-plane nerve fibers (iv). The straight colored lines indicate the fiber orientations known from anatomical brain structures (cc = corpus callosum, cr = corona radiata, cg = cingulum). The image on the right shows the strength of birefringence for a zoomed-in region of the corona radiata, measured with polarization microscopy \cite{MAxer2011_1,MAxer2011_2}. The fine yellow curves show the approximate pathways for different nerve fiber bundles, according to visible structures in the fiber architecture. \textbf{(b)} Measured scattering patterns and normalized (smoothed) polar integrals of the four regions (i)--(iv) indicated in (a). The non-dashed, colored lines indicate the positions of the scattering peaks, the dashed colored lines (in (i),(ii),(iv)) the in-plane orientation of the nerve fibers in the measured region. \textbf{(c)} Simulated scattering pattern and polar integral for nerve fibers with an out-of-plane angle of $\alpha = 60^{\circ}$ (adapted from \cite{menzel2020}, Supplementary Fig.\ S3, licensed under CC BY 4.0). The white circle indicates the area belonging to NA = 0.4.}
	\label{fig:scatterometry_vervet}
\end{figure}

\Cref{fig:scatterometry_vervet} shows the measured scattering patterns for different nerve fiber constellations in a vervet brain section (no.\ 458): parallel in-plane fibers in the \textit{corpus callosum} (i), crossing in-plane fibers in the \textit{corona radiata} ((ii),(iii)), and fibers pointing out of the section plane in the \textit{cingulum} (iv).

Just as for the model system of the crossing optic tracts, the region with parallel in-plane fibers yields two $180^{\circ}$-peaks (red lines in the polar integral) which are perpendicular to the nerve fibers in the corpus callosum (cc). 
The regions with crossing fibers yield two $180^{\circ}$-peak pairs (blue/green lines in the polar integrals) which are expected to be perpendicular to the respective fiber orientations (dashed blue/green lines in the scattering pattern). The two peak-pairs suggest that the measured regions contain two crossing fiber bundles. When studying the crossing fiber pathways in the corona radiata (cr) in more detail (see dashed yellow lines in the zoomed-in area in \cref{fig:scatterometry_vervet}(a)), it turns out that the fiber orientations determined from the scattering patterns (green/blue lines) correspond to the overall fiber orientations of several, intermingling crossing fiber bundles.
This shows that the measured scattering patterns reveal the overall fiber orientations not only for a simple model system of crossing optic tracts, but also in regions with more complicated, crossing fiber structures in whole brain sections.

\Cref{fig:scatterometry_vervet}(c) shows the simulated scattering pattern for a fiber bundle with an out-of-plane angle of $\alpha = 60^{\circ}$ (top image). 
While the light for in-plane fibers is scattered perpendicularly to the fiber direction (see \cref{fig:scatterometry_chiasm}(c)(i) for $\alpha = 0^{\circ}$), it is broadly scattered in the direction of the fibers for $\alpha = 60^{\circ}$ (here: along the x-direction; dashed magenta line), causing the two scattering peaks (non-dashed magenta lines) to move closer together. 
The measured scattering pattern of the out-of-plane fibers (iv) shows indeed a much broader scattering and more noise than for in-plane fibers (i), and the two peaks in the polar integral lie closer together ($150^{\circ}$ instead of $180^{\circ}$).
The middle position between the scattering peaks (dashed magenta line) corresponds to the (in-plane) fiber orientation of the measured region in the cingulum (cg).
This shows that also for out-of-plane fibers, the measured scattering pattern corresponds to the simulated scattering pattern. Note again that the scattering pattern was measured with NA = 0.4 and can only be compared to the inner part of the simulated scattering pattern (white circle), where the scattering peaks lie not as close together as when integrating over the full scattering pattern.

\subsection{Reconstruction of nerve fiber orientations}

In the previous sections, we have shown that the measured scattering patterns obtain information about the (in-plane) nerve fiber orientations in the measured tissue regions: The arithmetic mean values of the determined ($180^{\circ}$-)peak-pair positions in the (smoothed) polar integrals (vertical colored lines in \cref{fig:scatterometry_chiasm}(b) and \ref{fig:scatterometry_vervet}(b)) can be used to reconstruct the nerve fiber orientations in the brain section (straight colored lines in \cref{fig:scatterometry_chiasm}(a) and \ref{fig:scatterometry_vervet}(a)).

\begin{figure}[htbp]
	\includegraphics[width=0.85\textwidth]{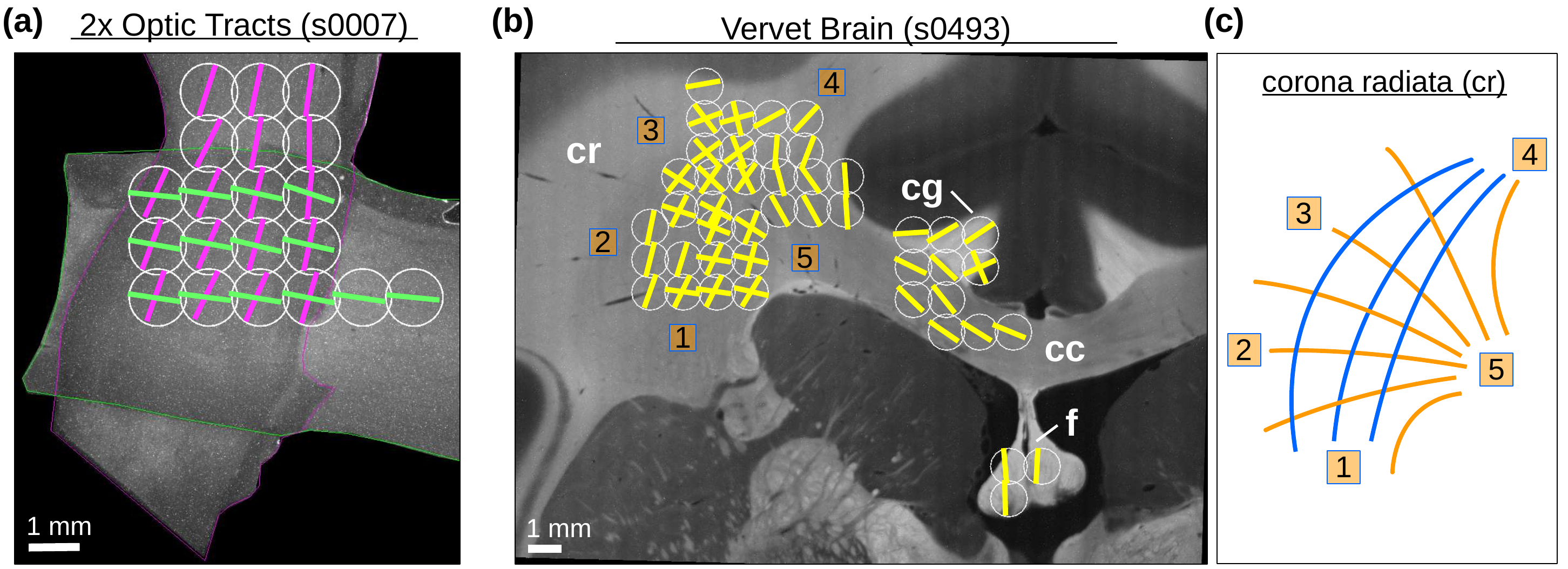}
	\caption{Reconstructed nerve fiber orientations for (a) two crossing sections of human optic tracts and (b) a coronal vervet monkey brain section. The images were obtained by dark-field microscopy; the sections of optic tracts in (a) were surrounded by a magenta/green outline for better reference. Different tissue regions were measured with scatterometry ($\varnothing = 1.12$\,mm, NA = 0.4,  $t =$ 30 ms), see white circles. The peak positions were determined from the smoothed polar integrals of the resulting scattering patterns, as shown in \cref{fig:scatterometry_chiasm,fig:scatterometry_vervet}. The (in-plane) fiber orientations were computed from the arithmetic mean values of the peak pair positions with approx.\ $180^{\circ}$ distance (cf.\ dashed green/blue lines in \cref{fig:scatterometry_vervet}(b)(ii)), and marked in the images by green, magenta, and yellow lines. \textbf{(c)} Sketch of crossing nerve fiber pathways in the corona radiata of the vervet brain section, known from polarization microscopy studies. (cr = corona radiata, cg = cingulum, cc = corpus callosum, f = fornix)}	
	\label{fig:fiber-orientations}
\end{figure}

To demonstrate the potential of this method, we reconstructed the nerve fiber orientations for two crossing sections of optic tracts (human chiasm, section no.\ 7) and a vervet brain section (no.\ 458) from the measured scattering patterns. \Cref{fig:fiber-orientations} shows the resulting fiber orientations (green/magenta/yellow lines) for the measured tissue regions (white circles).
The nerve fibers of the two crossing sections of optic tracts (in green/magenta) are clearly visible both within one section of the optic tract and in the crossing region.
The reconstructed nerve fiber orientations in the vervet brain section correspond to known anatomical fiber structures, both for in-plane fibers in the corpus callosum (cc) and for out-of-plane fibers in the cingulum (cg) and fornix (f). 
The sketch on the right-hand side illustrates the approximate pathways of the fiber bundles in the crossing region of the corona radiata (cr), known from polarization microscopy studies. The reconstructed fiber orientations (yellow lines) correspond very well to these pathways: We observe fibers running from [1] to [4] (corresponding to the blue fiber pathways), as well as fibers running from [5] to [1], [2], [3], and [4] (corresponding to the orange fiber pathways).


\section{Discussion and conclusion}

Nerve fiber crossings in the brain pose a major challenge for current neuroimaging techniques.
Simulation studies by Menzel \ea\ (2020a) \cite{menzel2020} suggest that the scattering patterns of light transmitted through brain tissue samples reveal valuable information about the tissue substructure like the individual orientations of in-plane crossing nerve fibers.
In this paper, we introduced a method based on coherent Fourier scatterometry that allows for the first time to measure these scattering patterns, validate the simulation approach, and reveal the orientations of crossing nerve fibers in unstained histological brain sections (monkey and human brain).

\subsection{Reconstruction of nerve fiber crossings from measured scattering patterns}

In contrast to the scattering measurement introduced in \cite{menzel2020}, our method allows to measure the \textit{full} scattering pattern (only limited by the numerical aperture of the objective lens). The measured scattering patterns can be used to reliably reconstruct the overall nerve fiber orientation in the measured tissue region with high angular precision ($< 1^{\circ}$). In regions with two or three crossing fiber bundles/layers, we separated the individual nerve fiber orientations (see \cref{fig:scatterometry_chiasm}). The smallest crossing angle between neighboring bundles that was considered here is about $25^{\circ}$ (see \cref{fig:pinhole-NA}). The minimum crossing angle that can still be resolved is determined by the width of the scattering peaks and the signal-to-noise ratio in the resulting polar integrals. The polar integrals in the blue box in \cref{fig:pinhole-NA} on the left suggest that it is not possible to distinguish nerve fiber bundles with much smaller crossing angles. This defines a lower bound for determining crossing angles in brain tissue using scattered light --- an important finding for further development of scattering measurements.
The nerve fiber orientations of crossing (in-plane) fibers were not only correctly determined for a model system of crossing fiber bundles (two and three crossing sections of human optic tracts, see \cref{fig:scatterometry_chiasm,fig:fiber-orientations}(a)). We could also demonstrate that the measured scattering patterns can be used to reliably determine more complex crossing fibers in whole brain sections, \eg\ in the corona radiata of a vervet monkey brain (see \cref{fig:scatterometry_vervet,fig:fiber-orientations}(b)).
We only investigated up to three crossing sections of optic tracts because a larger number of crossing fiber bundles is not likely to occur in whole brain tissue samples (for the investigated tissue regions of 0.1--1.1\,mm diameter), and a larger number of crossing sections would also increase the thickness of the sample and decrease the signal-to-noise ratio. In principle, the scatterometry measurement is able to resolve more than three different crossing fiber bundles, provided the crossing angles are sufficiently large.

The nerve fiber orientations were determined from the peak positions in the \textit{smoothed} polar integrals (see \cref{sec:evaluation-scatteringpattern}). However, our results suggest that the zigzag-structure in the \textit{non-smoothed} integrals also contains information about the substructure of the tissue (the zigzag-structure is time-independent and specific for each tissue region, see \cref{fig:noise}). Future studies should therefore consider the whole scattering pattern and original (non-smoothed) signals, and investigate how they can be used to obtain extra information about the tissue substructure. In addition, it would be interesting to use objective lenses with larger numerical apertures (even NA $>$ 1) and study if the measurements yield more information about the scattering properties of brain tissue.

\subsection{Validation of simulated scattering patterns}

In this paper, we provide for the first time a direct validation of the simulation approach by Menzel \ea\ (2020a) \cite{menzel2020}, who used finite-difference time-domain (FDTD) simulations to model light scattering in brain tissue.
Using a setup in the style of the simulations (\ie\ transmitting a coherent, non-focused laser beam with normal incidence through a brain section and measuring the distribution of the scattered light), we were able to measure scattering patterns for different brain tissue regions and compare them to the simulated scattering patterns in \cite{menzel2020}.

In contrast to the simulated scattering patterns, the measured scattering patterns are limited by the numerical aperture of the objective lens (here: NA $\leq 0.8$) so that they can only be compared to the inner part ($\theta \leq 53^{\circ}$) of the simulated scattering patterns. The simulated scattering patterns, on the other hand, have a limited resolution in $k$-space, \ie\ a limited number of scattering angles, because of limitations in computing time and largest possible sample size. Therefore, the polar integrals of the simulated scattering patterns were computed for NA = 1 and not for the numerical aperture of the imaging system (NA $\leq 0.8$). Nevertheless, we could show that the measured and simulated scattering patterns are very similar to each other --- both for in-plane (crossing) fibers and for out-of-plane fibers (see \cref{fig:scatterometry_vs_sim,fig:scatterometry_chiasm,fig:scatterometry_vervet}(c)). 

The simulated scattering patterns in \cite{menzel2020} were generated from volumes of $30 \times 30 \times 30$\,\um$^3$, while the measured scattering patterns were mostly obtained from 1.12\,mm large regions (defined by the laser beam diameter) in 30--60\,\um\ thin brain sections. However, we could show that the scattering patterns for beam diameters of 100\,\um\ yield similar features as for 1.12\,mm (see \cref{fig:pinhole-NA}). This is a similar order of magnitude as the simulated volume in \cite{menzel2020} and about the same size as the simulation volume ($128 \times 128 \times 60$\,\um$^3$) used in another publication \cite{reuter2019}, which yielded similar simulated scattering patterns.

Hence, our measurement results can serve as direct validation of the simulation approach in \cite{menzel2020}. As the measured scattering patterns correspond very well to the simulated ones, the FDTD simulation framework --- including the simplified model for the nerve fiber structure and the optics of the imaging system --- can be used to make valid predictions for the scattering behavior of fibrous brain tissue samples. 
As mentioned in \cite{menzel2020}, the developed simulation model can easily be adapted to other imaging systems and (non-biological) tissue samples with similar fibrous structures, allowing applications beyond neuroscience.

\subsection{Variations of measurement setup and sample}

So far, coherent Fourier scatterometry (CFS) has been applied to study scattering in non-biological, periodic samples \cite{kumar2014,gawhary2011}. To measure scattering patterns in brain tissue, we slightly modified the standard CFS setup in the style of the simulations, using \textit{non}-focused laser light in \textit{transmission} mode. Preliminary simulation studies have shown that the characteristic scattering patterns are only observed in the transmitted and not in the reflected light. As the measurement needs to be performed in transmission mode, in-vivo measurements \eg\ through a cranial window would not be possible. Furthermore, the tissue sample needs to be transparent enough to shine laser light through. For brain tissue with myelinated nerve fibers, this puts an upper limit on the maximum possible sample thickness. Thicker samples increase the overall scattering, reduce the transmitted light intensity, and lead to a lower signal-to-noise ratio (cf.\ \cref{fig:pinhole-NA}, on the right). For tissue samples with two crossing sections of optic tracts with $2 \times 60$\,\um\ thickness (s0007), we already have to wait several days to reduce the overall scattering and be able to image the crossing region.

A comparison of diffusion MRI images of brains before fixation with polarization microscopy measurements of several consecutive brain sections (and anatomical knowledge) showed that the overall nerve fiber architecture and anatomical structures do not change much during the preparation of the brain tissue samples (see \cref{sec:preparation}). The embedding in an aqueous solution (glycerin) prevents the tissue from dehydration and from structural changes. The glycerin solution is used as embedding medium because it is used as cryoprotectant, prevents the development of crystals, and yields good birefringence contrasts for polarization microscopy as well as a good scattering contrast. The simulations in \cite{menzel2020} have shown that the scattering patterns are mostly determined by the geometry and refractive indices of the sample. The simulation model assumes different refractive indices for axon (n = 1.35), myelin (n = 1.47), and surrounding glycerin solution (n = 1.37). 
As both the overall fiber geometry and the refractive indices are similar for fixed and non-fixed brain tissue, we would expect similar measurement results for non-fixed brain tissue samples with myelinated nerve fibers.

The simulations have shown that the contrast of the measured scattering patterns depends very much on the refractive index differences of the individual tissue components. As discussed in \cref{sec:pinhole-size}, the overall scattering of the sample decreases with increasing time after tissue embedding because the glycerin solution dries out and changes the refractive index differences between neighboring tissue components/layers. To obtain scattering patterns with high contrast, the refractive index of the embedding medium should differ from the refractive index of the fibers. In regions with unmyelinated nerve fibers with a low refractive index (axon: n = 1.35), for example, another embedding medium with a much lower/higher refractive index should be chosen.

As the scattering patterns are mostly determined by the fiber geometry and the refractive index differences of the components, the scatterometry measurement can also be used to study fiber crossings in other (also non-biological) samples with similar fibrous structures, \eg\ muscle fibers, collagen structures (in the sclera or lamina cribrosa of the eye \cite{jan2015}), or artificial fibers with comparable dimensions.

\subsection{Limitations and alternative approach}

Although the scatterometry measurement yields highly-resolved scattering patterns and can be used as validation of the simulated results, it has several limitations: 
First, the scatterometry measurement does not allow to exactly locate the measured tissue region --- this needs to be done in additional measurements (see \cref{sec:localization-laser}).
Since the reconstruction of the fiber orientations requires a separate measurement for each tissue region (with diameters of 0.1--1.1\,mm), our method can only be applied to study a certain number of small tissue regions (cf.\ \cref{fig:fiber-orientations}). Rasterizing a whole (human) brain section is not feasible.
Second, the sample needs to be illuminated by an approximately plane wave. Therefore, the laser beam diameter needs to be much larger than the wavelength ($\geq 100$\,\um), which limits the spatial resolution. In very inhomogeneous, crossing brain regions, the measured scattering patterns are therefore influenced by many different fiber orientations and cannot reveal the course of individual nerve fibers. The resolution achieved by the scatterometry measurement is comparable to the one achieved by post-mortem diffusion MRI. However, the here presented method can be applied to (non-)biological samples with similar fibrous structures in order to resolve crossing fibers in bulk tissue that can otherwise not be resolved (if dMRI cannot be applied or is not available).

Our scatterometry measurement yields the \textit{full} scattering pattern for a \textit{single} brain tissue region ($\varnothing = 0.1$--$1.12$\,mm).
To study crossing nerve fibers in whole brain sections, Menzel \ea\ \cite{menzel2020} introduced a scattering measurement with oblique illumination. The latter technique measures only a \textit{limited} number of scattering angles in the scattering pattern, but for \textit{all} image pixels at once (\ie\ with micrometer resolution).

Revealing individual nerve fiber orientations in regions with crossing fibers is a major challenge for many neuroimaging techniques, and highly significant when it comes to a correct tractography of nerve fiber pathways in the brain. A major result of our study is that we could for the first time measure the scattering patterns of brain tissue and validate the simulated scattering patterns in \cite{menzel2020}. This allows to use the simulated patterns as a reference to further improve the scattering measurement with oblique illumination (\eg\ by selecting the optimum scattering angles for the measurement).
In the longer term, this will allow for a more detailed reconstruction of (crossing) nerve fiber pathways in the brain and for a better understanding of neurodegenerative diseases.


\section*{Author contributions}
M.M.\ developed the concept and design of the study, performed the scatterometry measurements (with assistance from S.P.), analyzed the data, and wrote the manuscript (with input from S.P.). S.P.\ designed and optimized the set-up for the scatterometry measurements and assisted in technical questions. Both authors approved the final version of the manuscript.

\section*{Acknowledgments}
We would like to thank Markus Cremer, Patrick Nysten, and Steffen Werner for the preparation of the histological brain sections, David Gräßel for providing the transmittance and dark-field microscopy images, Thim Zuidwijk for assistance with the digital microscope measurements, and Jan André Reuter for the smoothing of the line profiles. Furthermore, we thank Katrin Amunts, Markus Axer, Karl Zilles, and Roger Woods for collaboration in the vervet brain project, and Roxana Kooijmans and the Netherlands Brain Bank for providing the human optic chiasm.
M.M.\ thanks Markus Axer and Katrin Amunts for useful discussion and their continuous support, and the Department of Imaging Physics, TU Delft, for hosting her and the fruitful collaboration.
This project has received funding from the European Union's Horizon 2020 Framework Programme for Research and Innovation under the Specific Grant Agreement No.\ 785907 and 945539 (``Human Brain Project'' SGA2 and SGA3) and the National Institutes of Health (Grant Agreement No.\ 4R01MH092311).
M.M.\ received funding from the Helmholtz Association port-folio theme (Supercomputing and Modeling for the Human Brain) and the Helmholtz Doctoral Prize 2019.

\appendix

\section*{Appendix}


\section{Preparation of brain sections}
\label{sec:preparation}

The measurements were performed on sections of a vervet monkey brain (African green monkey: \textit{Chlorocebus aethiops sabaeus}, male, between 1--2 years old) and a human optic chiasm (female, 74 years old), both healthy subjects. 
All animal procedures have been approved by the institutional animal welfare committee at Forschungszentrum Jülich GmbH, Germany, and are in accordance with National Institutes of Health guidelines for the use and care of laboratory animals. The human brain was acquired from the Netherlands Brain Bank, in the Netherlands Institute for Neuroscience, Amsterdam. A written informed consent of the subject is available. 

The brains were obtained within 24 hours after death and fixed in a buffered solution of 4\% formaldehyde for several weeks. Subsequently, the brains were immersed in a solution of 10\% glycerin and then in a solution of 20\% glycerin for several days to avoid the development of ice crystals. To facilitate the penetration of glycerin into the cells, 2\% dimethyl sulfoxide and 4\% formaldehyde were added. The brains were deeply frozen, and cut with a crystat microtome (\textit{Polycut CM 3500}, \textit{Leica Microsystems}, Germany) into histological sections.
The vervet brain was cut along the coronal plane into sections of 60\,\um, and sections no.\ 458 and 493 were selected for the measurements. The chiasm (including at least 1\,cm of the optic tracts and the optic nerves) was cut along the fiber tracts of the visual pathway into sections of 30\,\um\ and 60\,\um, and section no.\ 7 (60\,\um) and sections no.\ 15, 32, 33, 36 (30\,\um) were selected for the measurements.
In order to obtain well-defined samples with two or three crossing fiber bundles/layers, the selected sections of the optic chiasm were cut into two parts at the median line, and the sections of the optic tracts (left and right) were manually placed on top of each other with different crossing angles (cf.\ \cref{fig:scatterometry_samples}(b)). The brain samples were mounted on glass slides, embedded in a solution of 20\% glycerin, covered by a cover glass, sealed with lacquer, and weighted for several hours to prevent the development of air bubbles, before measuring the samples. \Cref{fig:scatterometry_samples}(c) shows the dark-field microscopy measurements of all investigated brain tissue samples.


\section{Measurement setup}
\label{sec:scatterometry}

\begin{figure}[htbp]
	\centering\includegraphics[width=0.7\textwidth]{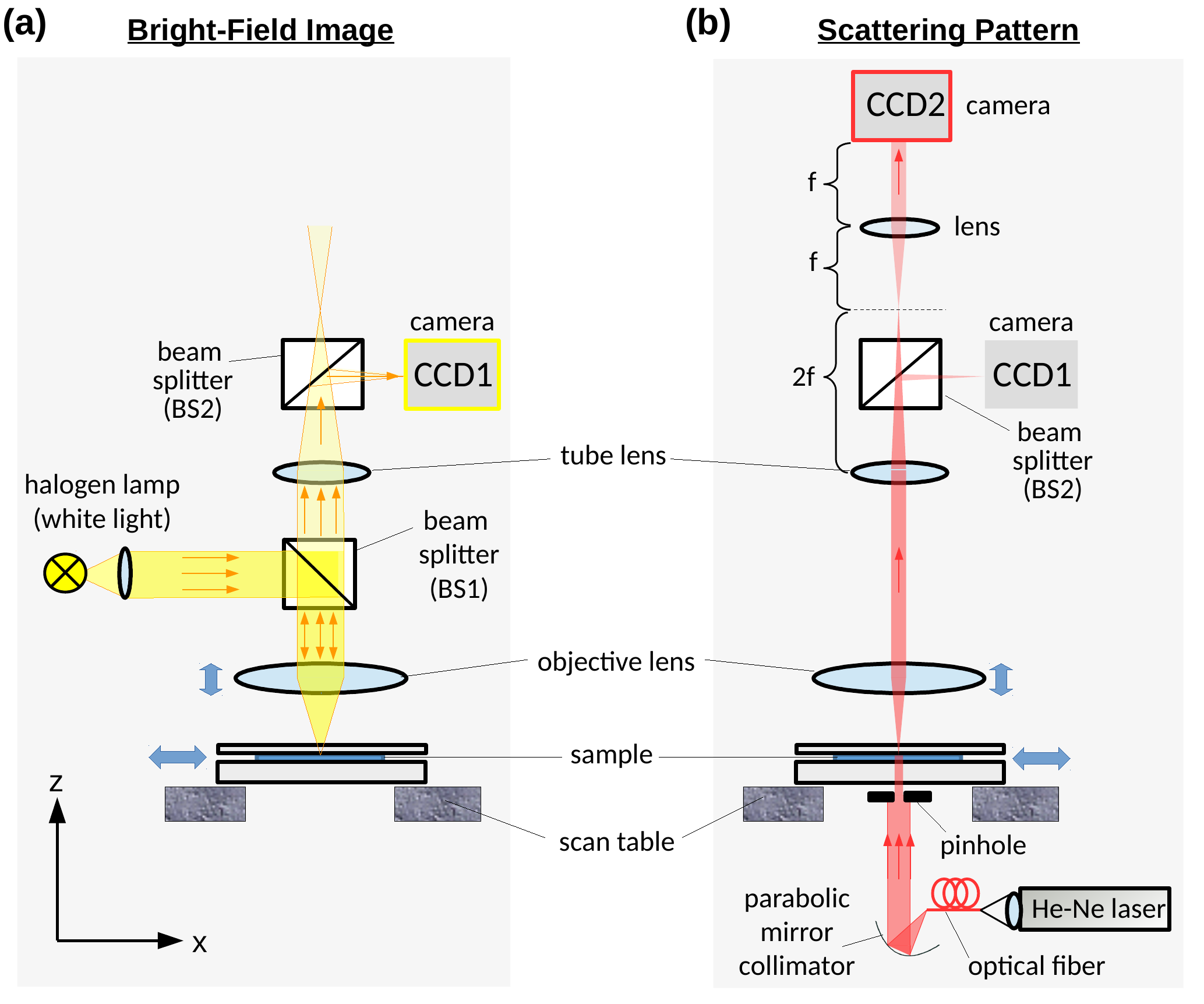}
	\caption{Schematic drawing of the setup used for the scatterometry measurements: \textbf{(a)} Setup used to record a bright-field image of the sample plane. \textbf{(b)} Setup used to record the Fourier transform of the image plane (scattering pattern). (CCD = charge-coupled device)}
	\label{fig:scatterometry-setup}
\end{figure}

The measurement setup (see Fig.~\ref{fig:scatterometry-setup}) is similar to the one used by Kumar \ea\ (2014) \cite{kumar2014} to perform coherent Fourier scatterometry (CSF) on printed gratings. While those authors focused the laser light onto the sample and measured the reflected light, the collimated laser light is here transmitted through the sample without any focusing (normally incident light, cf.\ Fig.~\ref{fig:scatterometry-setup}(b)).

In order to measure different positions of the specimen, the sample was placed on a scan table which can be moved in the x- and y-direction using micrometer screws.

Figure~\ref{fig:scatterometry-setup}(a) illustrates how the \textit{bright-field images} of the sample (cf.\ \cref{fig:scat-point}(b)) were recorded:
Light from a white light source (halogen lamp) is directed by a beam splitter (BS1) through an objective lens onto the sample. The light reflected by the sample is collected by the objective lens, passes through the beam splitter and a tube lens (focal length: $2f = 16$\,cm) which focuses the light onto a camera (CCD1), generating an image of the sample plane.

Figure~\ref{fig:scatterometry-setup}(b) illustrates how the \textit{scattering patterns} of the sample were recorded:
Light from a helium-neon laser ($\lambda = 632.8$\,nm) is guided through a single mode optical fiber and collimated by a parabolic mirror so that the sample is illuminated by normally incident, coherent light from below.  
A pinhole (with a diameter of 0.1\,mm or 1.12\,mm) is placed right below the sample to define the area of illumination.
The scattered laser light behind the sample is collected by the objective lens and focused by the tube lens onto the camera CCD1 (image of sample plane). Another beam splitter (BS2) guides part of the laser light to another lens ($f = 8$\,cm) which makes a Fourier transform of the image plane at the camera CCD2.
During data acquisition, the first beam splitter (BS1) is removed so that only the light from the laser beam is recorded by the two CCD cameras.

The white light illumination (Fig.~\ref{fig:scatterometry-setup}(a)) was realized by a scanning near-field optical microscope operated in reflection mode (\textit{WITec AlphaSNOM}, manufactured by \textit{Wissenschaftliche Instrumente und Technologie GmbH}, Germany). 
The objective lenses used for the measurements are cover-glass corrected and have different numerical apertures (NA):
\begin{itemize}
	\item Single lens: NA $=$ 0.14, focal length $= 3.5$\,cm, diameter $=$ 10\,mm.
	\item \textit{Nikon CFI Achro 60X}: NA $=$ 0.8, $60\times$ magnification, working distance $= 3$\,mm, cover glass thickness $= 0.17$\,mm, chromatic aberration free infinity (CFI),
	\item \textit{Nikon CFI Achro LWD 20X}: NA $=$ 0.4, $20\times$ magnification, working distance $= 3.8$\,mm, cover glass thickness $= 0.17$\,mm, chromatic aberration free infinity (CFI),
\end{itemize}

The first camera (CCD1) is part of the \textit{WITec AlphaSNOM} and records $1024 \times 768$ pixels with a size of $0.6 \times 0.6$\,\textmu m$^2$ per pixel.
The second camera (CCD2) is a \textit{PROSILICA GC1290}, manufactured by \textit{Allied Vision Technologies GmbH}. This CCD camera is a 12 bit camera with a resolution of $1280 \times 960$ pixels, a pixel size of $3.75 \times 3.75$\,\textmu m$^2$, and a sensing area of $4.8 \times 3.6$\,mm$^2$.


\section{Computation of azimuthal and polar integrals}
\label{sec:evaluation-scatteringpattern}

\begin{figure}[htbp]
	\centering\includegraphics[width=0.8\textwidth]{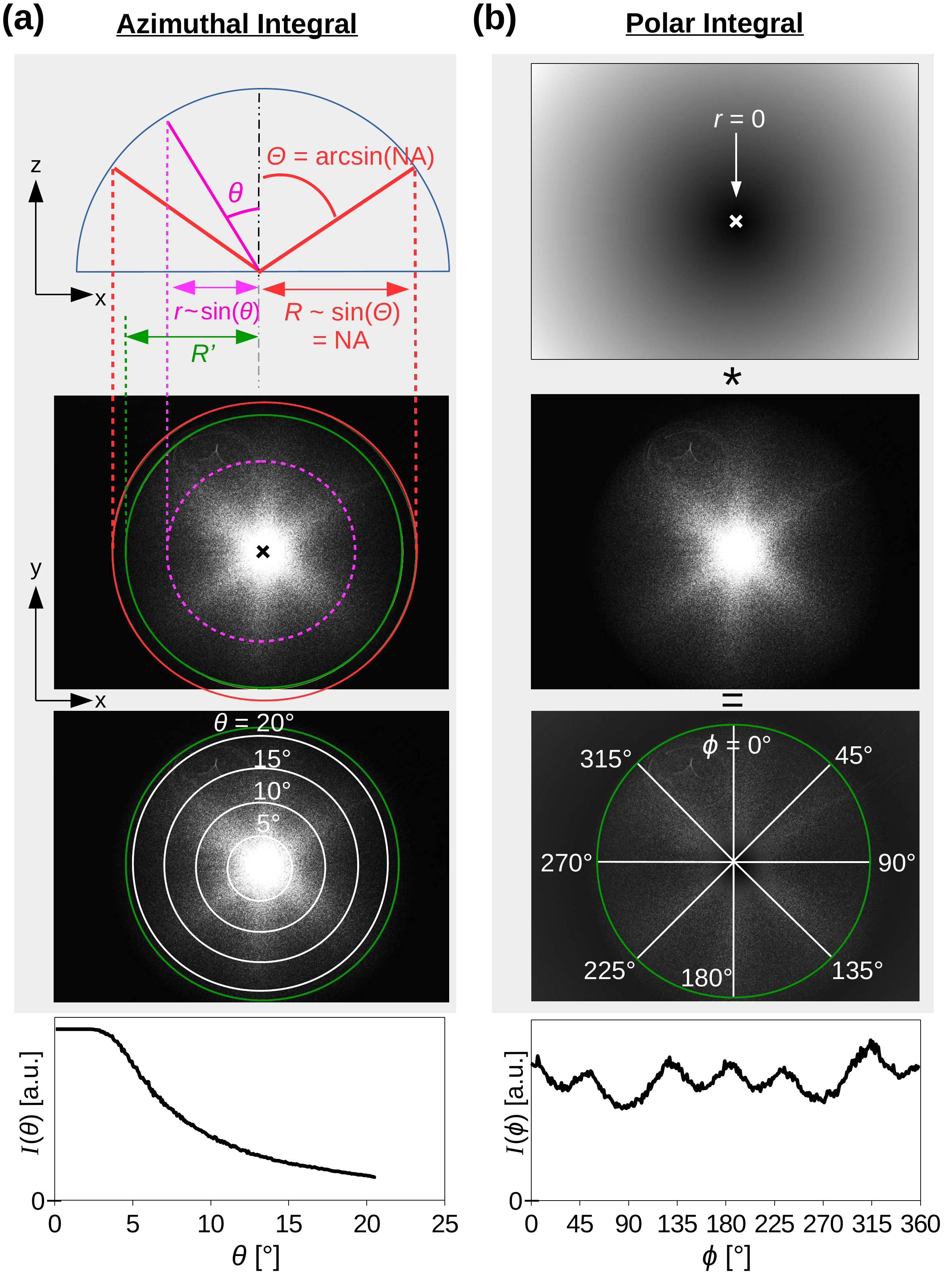}
	\caption{Evaluation of scattering patterns, shown exemplary for a region with three crossing sections of optic tracts (see \cref{fig:scatterometry_chiasm}(iv)): \textbf{(a)} azimuthal integral, \textbf{(b)} polar integral. The red circle with radius $R$ corresponds to the maximum scattering angle $\varTheta$ that is collected by the objective lens with numerical aperture NA. The green circle with radius $R'$ is the largest possible circle that can still be evaluated. To compute the azimuthal integral for a certain polar angle $\theta$, the intensity values of the scattering pattern are integrated over a concentric circle with radius $\sin(\theta)$ (see lower image in (a)). To compute the polar integral for a certain azimuthal angle $\phi$, the intensity values of the scattering pattern are multiplied by the distance $r$ to the center (upper image in (b)) and integrated from the center ($r=0$) to the outer circle ($r=R'$) for the corresponding $\phi$ (see lower image in (b)). The graphs show the azimuthal integral $I(\theta)$ and the polar integral $I(\phi)$, computed with \cref{eq:azimuthal-int,eq:polar-int}.}
	\label{fig:scatterometry-evaluation}
\end{figure}

The scattering patterns show the Fourier transform of the image plane, \ie\ the distribution of scattered light in a hemisphere behind the sample projected onto the sample plane (cf.\ \cref{fig:scatterometry_vs_sim}(c)), and are limited by the numerical aperture (NA) of the objective lens. 
To quantify the scattering patterns, the azimuthal integral $I(\theta)$ and the polar integral $I(\phi)$ were computed for each scattering pattern:
\begin{align}
I(\theta) &= I(r) = \int_0^{2\pi} I(r,\phi) \, \mathrm{d}\phi,  \label{eq:azimuthal-int} \\
I(\phi) &= \int_0^{R'} I(r,\phi) \, r \, \mathrm{d}r, \label{eq:polar-int}
\end{align}
where $r = \sin(\theta)$ is the distance to the center of the scattering pattern, and $\theta$ and $\phi$ are the polar and azimuthal angles in spherical coordinates as defined in Fig.\ \ref{fig:scatterometry-evaluation}. Note that $\phi=0^{\circ}$ was here chosen to be aligned with the $y$-axis with \textit{clock-wise} rotation.

Figure \ref{fig:scatterometry-evaluation} illustrates how the azimuthal and polar integrals were computed:
To identify the center of the scattering pattern ($\theta=0^{\circ}$, $r=0$), the contrast of the recorded image was enhanced so that the border between scattering pattern and background becomes visible (red circle). The intensity values were evaluated within the largest possible circle around the center that does not extend beyond the edges of the image (green circle). To compute the azimuthal integral $I(\theta)$, the intensity values were integrated over concentric circles with radius $r = \sin(\theta)$ and plotted against $\theta$ (see Fig.\ \ref{fig:scatterometry-evaluation}(a)). To compute the polar integral $I(\phi)$, the intensity values for each image pixel were multiplied by the distance to the center $r$, integrated from the center $r=0$ to the outer (green) circle for a given azimuthal angle $\phi$, and plotted against $\phi$ (see Fig.\ \ref{fig:scatterometry-evaluation}(b)).

To facilitate the determination of peak positions in the resulting polar integrals, the line profiles were smoothed out using a Savitzky-Golay filter with 45 sampling points and a second order polynomial \cite{savitsky1964}.


\section{Polarization effects}
\label{sec:polarization}

\begin{figure}[htbp]
	\centering\includegraphics[width=0.5\textwidth]{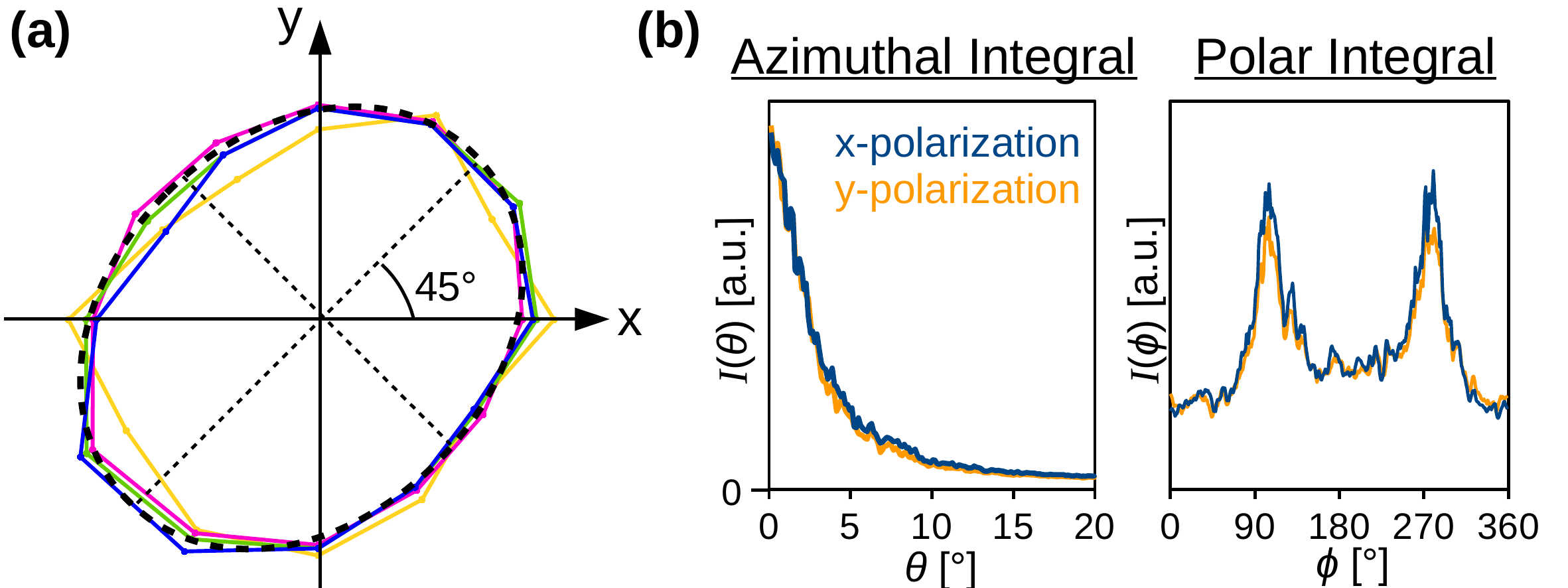}
	\caption{Study of polarization effects. \textbf{(a)} Polar plot of the average transmitted light intensity for different rotation angles \{$0^{\circ}$, $30^{\circ}$, $\dots$, $330^{\circ}$\} of a linear polarizer placed directly behind the 1.12\,mm pinhole. The measurement was repeated four times (colored curves) and fitted by an elliptical shape (black dashed curve). \textbf{(b)} Azimuthal and polar integrals of a scattering pattern measured with 1.12\,mm pinhole, NA = 0.4, and linearly polarized light (blue: polarization along the x-axis, orange: polarization along the y-axis).}
	\label{fig:scatterometry-polarization}
\end{figure}

To determine the polarization of the incident laser light, a linear polarizer was directly placed behind the 1.12\,mm pinhole (with the transmission axis aligned along the y-axis of the specimen stage) and rotated in steps of $30^{\circ}$ in counter-clockwise direction. For each rotation angle, the intensity of the pixel with the maximum intensity value was measured (average over time). To avoid saturation, an exposure time of 150\,\textmu s was used and a filter was directly placed behind the light source to reduce intensity. The measurement was performed four times to obtain statistics. \Cref{fig:scatterometry-polarization}(a) shows the measured intensities for the different rotation angles in a polar plot. 
The four different measurements (colored curves) can be fitted by an ellipse (dashed black curve). 
Hence, the laser light is elliptically polarized with the major axis rotated by $+45^{\circ}$ with respect to the x-axis of the sample and an eccentricity of 0.6.

Polarization-dependent light scattering in comparable brain sections was shown to be small \cite{menzel2019,menzel2017}. In fact, the measured scattering patterns and azimuthal/polar integrals do not depend very much on the polarization of the incident light. \Cref{fig:scatterometry-polarization}(b) shows the azimuthal and polar integrals computed from a scattering pattern for light polarized along the x-direction (blue) and along the y-direction (orange). The curves are almost identical. Polarization effects were therefore not considered in this study.

\section{Investigated brain sections and measured brain regions}
\label{sec:scatterometry-samples}

\begin{figure}[h!]
	\includegraphics[width=0.9\textwidth]{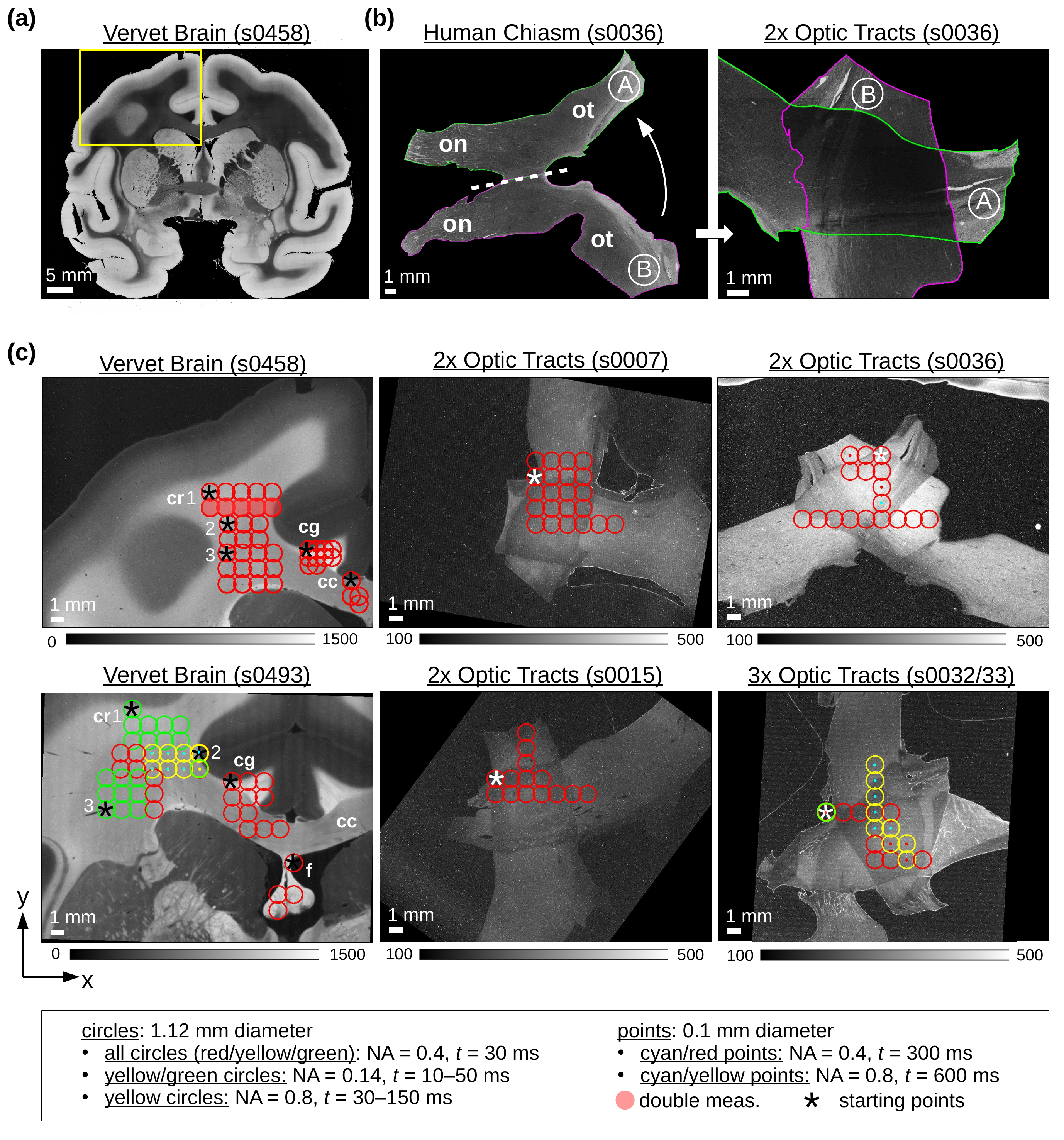}
	\caption{Investigated brain tissue samples. \textbf{(a)} Transmittance image of the coronal vervet monkey brain section (no.\ 458). The corresponding region shown in (c) is marked by a yellow rectangle. \textbf{(b)} Transmittance image of the human optic chiasm (section no.\ 36) for the whole section (left) and the two sections of the optic tracts crossing each other (right). (ot = optic tract, on = optic nerve) \textbf{(c)} Dark-field microscopy images of all investigated brain tissue samples. The colored circles and points show the tissue regions that were measured with scatterometry (using different pinhole diameters, numerical apertures (NA), and exposure times ($t$), refer to legend). The asterisk ($\ast$) marks the starting point of the measurement from which the sample (laser point position) was moved in steps of 0.5\,mm and 1\,mm (cf.\ \cref{fig:scat-point}). (cr = corona radiata, cg = cingulum, cc = corpus callosum, f = fornix)}
	\label{fig:scatterometry_samples}
\end{figure}

\Cref{fig:scatterometry_samples} shows all investigated brain tissue samples and measured tissue regions (for different beam diameters, numerical apertures, and exposure times). 
\Cref{tab:table} lists the sample properties and measurement parameters in more detail.
For all measured tissue regions, the scattering patterns and corresponding line profiles (azimuthal integrals and normalized polar integrals with smoothed curves) can be found in Dataset 1 (Ref.\ \cite{dataset}). The file names contain (separated by underlines): sample name (Vervet/OpticTracts), section number (s0007--s0493), beam diameter (100\,\um, 1120\,\um), numerical aperture (NA = \{0.14, 0.4, 0.8\}), exposure time (10, 30, 50, 150, 300, 600\,ms), brain region (cr1, cr2, cr3, cg, cc, f), and x/y-coordinates. Double measurements are marked by an additional underline. For example, measurement data called \textit{``Vervet\_s0493\_1120um\_NA-0,4\_30ms\_cr2\_x-3mm\_y+0mm''} belongs to the upper left yellow circle in the vervet brain section no.\ 493 (corona radiata) in \cref{fig:scatterometry_samples}.
\begin{table}[h]
	\normalsize
	\begin{center}
		\begin{tabular}{| c | c || c | c | c | c | c | c |}
			\hline
			\textbf{sample}			& \textbf{section}	& \textbf{$T$ [\um]} & \textbf{dark-field [d]} & \textbf{scatt. [d]} & \textbf{$\varnothing$ [mm]}	& \textbf{NA}	& \textbf{$t$ [ms]} 	\\ \hline\hline
			\multirow{5}{*}{\shortstack[|c|]{\textbf{Vervet Brain} \\[0.8mm] (coronal)}} 				
			& \textbf{458}							& 60 					& 13					& 22						& 1.12				& 0.4	& 30 \\ \cline{2-8}		
			& \multirow{4}{*}{\textbf{493}} 		& \multirow{4}{*}{60} 	& \multirow{4}{*}{9}	& 43 						& 1.12 				& 0.4	& 30 \\ \cline{5-8}
			& 	 									& 						& 						& \multirow{2}{*}{44}		& 1.12 				& 0.8	& 30, 50, 150 	\\ \cline{6-8}
			&										& 						& 						& 							& 0.10 				& 0.8	& 600 		\\ \cline{5-8}
			&										& 						& 						& 45						& 1.12 				& 0.14	& 10, 30, 50 \\ \hline\hline
			\multirow{9}{*}{\shortstack[|c|]{\textbf{Optic Tracts} \\[0.8mm] (human chiasm)}} 			
			& \textbf{7}		 					& 60				& 275						& 293						& 1.12				& 0.4	& 30		\\ \cline{2-8}
			& \textbf{15}							& 30				& 275						& 293						& 1.12				& 0.4	& 30		\\ \cline{2-8}
			& \multirow{3}{*}{\textbf{36}} 			& \multirow{3}{*}{30}	& \multirow{3}{*}{9}	& 43						& 1.12				& 0.4	& 30		\\ \cline{5-8}
			& 										& 					& 							& 44						& 0.10				& 0.4	& 300		\\ \cline{5-8}
			& 										& 					& 							& 44						& 0.10				& 0.8	& 600		\\ \cline{2-8}
			& \multirow{5}{*}{\textbf{32/33}}		& \multirow{5}{*}{30}	& \multirow{5}{*}{112}	& 126						& 1.12				& 0.4	& 30		\\ \cline{5-8}
			& 										& 					& 							& 127						& 0.10				& 0.4	& 300		\\ \cline{5-8}
			& 										& 					& 							& \multirow{3}{*}{128}		& \multirow{2}{*}{1.12}	& 0.14	& 10	\\ \cline{7-8}
			& 										& 					& 							& 							& 					& 0.8	& 100		\\ \cline{6-8}
			& 										& 					& 							& 				   			& 0.10				& 0.8	& 600		\\ \hline
		\end{tabular}
	\end{center}
	\caption{List of sample properties and measurement parameters for the investigated brain tissue samples: sample, section number, section thickness ($T$), dates of dark-field microscopy and scatterometry (scatt.) measurements [in days after tissue embedding], pinhole diameter ($\varnothing$), numerical aperture (NA), and exposure time ($t$) used for the scatterometry measurements.}
	\label{tab:table}
\end{table}

\cleardoublepage
\bibliography{BIBLIOGRAPHY}

\begin{thebibliography}{23}%
\makeatletter
\providecommand \@ifxundefined [1]{%
 \@ifx{#1\undefined}
}%
\providecommand \@ifnum [1]{%
 \ifnum #1\expandafter \@firstoftwo
 \else \expandafter \@secondoftwo
 \fi
}%
\providecommand \@ifx [1]{%
 \ifx #1\expandafter \@firstoftwo
 \else \expandafter \@secondoftwo
 \fi
}%
\providecommand \natexlab [1]{#1}%
\providecommand \enquote  [1]{``#1''}%
\providecommand \bibnamefont  [1]{#1}%
\providecommand \bibfnamefont [1]{#1}%
\providecommand \citenamefont [1]{#1}%
\providecommand \href@noop [0]{\@secondoftwo}%
\providecommand \href [0]{\begingroup \@sanitize@url \@href}%
\providecommand \@href[1]{\@@startlink{#1}\@@href}%
\providecommand \@@href[1]{\endgroup#1\@@endlink}%
\providecommand \@sanitize@url [0]{\catcode `\\12\catcode `\$12\catcode
  `\&12\catcode `\#12\catcode `\^12\catcode `\_12\catcode `\%12\relax}%
\providecommand \@@startlink[1]{}%
\providecommand \@@endlink[0]{}%
\providecommand \url  [0]{\begingroup\@sanitize@url \@url }%
\providecommand \@url [1]{\endgroup\@href {#1}{\urlprefix }}%
\providecommand \urlprefix  [0]{URL }%
\providecommand \Eprint [0]{\href }%
\providecommand \doibase [0]{http://dx.doi.org/}%
\providecommand \selectlanguage [0]{\@gobble}%
\providecommand \bibinfo  [0]{\@secondoftwo}%
\providecommand \bibfield  [0]{\@secondoftwo}%
\providecommand \translation [1]{[#1]}%
\providecommand \BibitemOpen [0]{}%
\providecommand \bibitemStop [0]{}%
\providecommand \bibitemNoStop [0]{.\EOS\space}%
\providecommand \EOS [0]{\spacefactor3000\relax}%
\providecommand \BibitemShut  [1]{\csname bibitem#1\endcsname}%
\let\auto@bib@innerbib\@empty
\bibitem [{\citenamefont {Herculano-Houzel}(2009)}]{herculano2009}%
  \BibitemOpen
  \bibfield  {author} {\bibinfo {author} {\bibfnamefont {S.}~\bibnamefont
  {Herculano-Houzel}},\ }\href {\doibase 10.3389/neuro.09.031.2009} {\bibfield
  {journal} {\bibinfo  {journal} {Frontiers in Human Neuroscience}\ }\textbf
  {\bibinfo {volume} {3}},\ \bibinfo {pages} {1} (\bibinfo {year}
  {2009})}\BibitemShut {NoStop}%
\bibitem [{\citenamefont {Shi}\ and\ \citenamefont {Toga}(2017)}]{shi2017}%
  \BibitemOpen
  \bibfield  {author} {\bibinfo {author} {\bibfnamefont {Y.}~\bibnamefont
  {Shi}}\ and\ \bibinfo {author} {\bibfnamefont {A.~W.}\ \bibnamefont {Toga}},\
  }\href {\doibase 10.1038/mp.2017.92} {\bibfield  {journal} {\bibinfo
  {journal} {Mol. Psychiatry}\ }\textbf {\bibinfo {volume} {22}},\ \bibinfo
  {pages} {1230} (\bibinfo {year} {2017})}\BibitemShut {NoStop}%
\bibitem [{\citenamefont {Maier-Hein}\ \emph {et~al.}(2017)\citenamefont
  {Maier-Hein}, \citenamefont {Neher}, \citenamefont {Houde}, \citenamefont
  {C\^{o}t\'{e}}, \citenamefont {Garyfallidis}, \citenamefont {Zhong},\ and\
  \citenamefont {Chamberland}}]{maierhein2017}%
  \BibitemOpen
  \bibfield  {author} {\bibinfo {author} {\bibfnamefont {K.~H.}\ \bibnamefont
  {Maier-Hein}}, \bibinfo {author} {\bibfnamefont {P.~F.}\ \bibnamefont
  {Neher}}, \bibinfo {author} {\bibfnamefont {J.-C.}\ \bibnamefont {Houde}},
  \bibinfo {author} {\bibfnamefont {M.-A.}\ \bibnamefont {C\^{o}t\'{e}}},
  \bibinfo {author} {\bibfnamefont {E.}~\bibnamefont {Garyfallidis}}, \bibinfo
  {author} {\bibfnamefont {J.}~\bibnamefont {Zhong}}, \ and\ \bibinfo {author}
  {\bibfnamefont {M.}~\bibnamefont {Chamberland}},\ }\href {\doibase
  10.1038/s41467-017-01285-x} {\bibfield  {journal} {\bibinfo  {journal} {Nat.
  Comm.}\ }\textbf {\bibinfo {volume} {8}},\ \bibinfo {pages} {1349} (\bibinfo
  {year} {2017})}\BibitemShut {NoStop}%
\bibitem [{\citenamefont {Axer}\ \emph
  {et~al.}(2011{\natexlab{a}})\citenamefont {Axer}, \citenamefont {Amunts},
  \citenamefont {Gr{\"a}ssel}, \citenamefont {Palm}, \citenamefont {Dammers},
  \citenamefont {Axer}, \citenamefont {Pietrzyk},\ and\ \citenamefont
  {Zilles}}]{MAxer2011_1}%
  \BibitemOpen
  \bibfield  {author} {\bibinfo {author} {\bibfnamefont {M.}~\bibnamefont
  {Axer}}, \bibinfo {author} {\bibfnamefont {K.}~\bibnamefont {Amunts}},
  \bibinfo {author} {\bibfnamefont {D.}~\bibnamefont {Gr{\"a}ssel}}, \bibinfo
  {author} {\bibfnamefont {C.}~\bibnamefont {Palm}}, \bibinfo {author}
  {\bibfnamefont {J.}~\bibnamefont {Dammers}}, \bibinfo {author} {\bibfnamefont
  {H.}~\bibnamefont {Axer}}, \bibinfo {author} {\bibfnamefont {U.}~\bibnamefont
  {Pietrzyk}}, \ and\ \bibinfo {author} {\bibfnamefont {K.}~\bibnamefont
  {Zilles}},\ }\href {\doibase 10.1016/j.neuroimage.2010.08.075} {\bibfield
  {journal} {\bibinfo  {journal} {Neuro{I}mage}\ }\textbf {\bibinfo {volume}
  {54}},\ \bibinfo {pages} {1091} (\bibinfo {year}
  {2011}{\natexlab{a}})}\BibitemShut {NoStop}%
\bibitem [{\citenamefont {Axer}\ \emph
  {et~al.}(2011{\natexlab{b}})\citenamefont {Axer}, \citenamefont
  {Gr{\"a}ssel}, \citenamefont {Kleiner}, \citenamefont {Dammers},
  \citenamefont {Dickscheid}, \citenamefont {Reckfort}, \citenamefont
  {H{\"u}tz}, \citenamefont {Eiben}, \citenamefont {Pietrzyk}, \citenamefont
  {Zilles},\ and\ \citenamefont {Amunts}}]{MAxer2011_2}%
  \BibitemOpen
  \bibfield  {author} {\bibinfo {author} {\bibfnamefont {M.}~\bibnamefont
  {Axer}}, \bibinfo {author} {\bibfnamefont {D.}~\bibnamefont {Gr{\"a}ssel}},
  \bibinfo {author} {\bibfnamefont {M.}~\bibnamefont {Kleiner}}, \bibinfo
  {author} {\bibfnamefont {J.}~\bibnamefont {Dammers}}, \bibinfo {author}
  {\bibfnamefont {T.}~\bibnamefont {Dickscheid}}, \bibinfo {author}
  {\bibfnamefont {J.}~\bibnamefont {Reckfort}}, \bibinfo {author}
  {\bibfnamefont {T.}~\bibnamefont {H{\"u}tz}}, \bibinfo {author}
  {\bibfnamefont {B.}~\bibnamefont {Eiben}}, \bibinfo {author} {\bibfnamefont
  {U.}~\bibnamefont {Pietrzyk}}, \bibinfo {author} {\bibfnamefont
  {K.}~\bibnamefont {Zilles}}, \ and\ \bibinfo {author} {\bibfnamefont
  {K.}~\bibnamefont {Amunts}},\ }\href {\doibase 10.3389/fninf.2011.00034}
  {\bibfield  {journal} {\bibinfo  {journal} {Front. Neuroinform.}\ }\textbf
  {\bibinfo {volume} {5}},\ \bibinfo {pages} {1} (\bibinfo {year}
  {2011}{\natexlab{b}})}\BibitemShut {NoStop}%
\bibitem [{\citenamefont {Reckfort}\ \emph {et~al.}(2015)\citenamefont
  {Reckfort}, \citenamefont {Wiese}, \citenamefont {Pietrzyk}, \citenamefont
  {Zilles}, \citenamefont {Amunts},\ and\ \citenamefont {Axer}}]{reckfort2015}%
  \BibitemOpen
  \bibfield  {author} {\bibinfo {author} {\bibfnamefont {J.}~\bibnamefont
  {Reckfort}}, \bibinfo {author} {\bibfnamefont {H.}~\bibnamefont {Wiese}},
  \bibinfo {author} {\bibfnamefont {U.}~\bibnamefont {Pietrzyk}}, \bibinfo
  {author} {\bibfnamefont {K.}~\bibnamefont {Zilles}}, \bibinfo {author}
  {\bibfnamefont {K.}~\bibnamefont {Amunts}}, \ and\ \bibinfo {author}
  {\bibfnamefont {M.}~\bibnamefont {Axer}},\ }\href {\doibase
  10.3389/fnana.2015.00118} {\bibfield  {journal} {\bibinfo  {journal} {Front.
  Neuroanat.}\ }\textbf {\bibinfo {volume} {9}},\ \bibinfo {pages} {1}
  (\bibinfo {year} {2015})}\BibitemShut {NoStop}%
\bibitem [{\citenamefont {Dohmen}\ \emph {et~al.}(2015)\citenamefont {Dohmen},
  \citenamefont {Menzel}, \citenamefont {Wiese}, \citenamefont {Reckfort},
  \citenamefont {Hanke}, \citenamefont {Pietrzyk}, \citenamefont {Zilles},
  \citenamefont {Amunts},\ and\ \citenamefont {Axer}}]{dohmen2015}%
  \BibitemOpen
  \bibfield  {author} {\bibinfo {author} {\bibfnamefont {M.}~\bibnamefont
  {Dohmen}}, \bibinfo {author} {\bibfnamefont {M.}~\bibnamefont {Menzel}},
  \bibinfo {author} {\bibfnamefont {H.}~\bibnamefont {Wiese}}, \bibinfo
  {author} {\bibfnamefont {J.}~\bibnamefont {Reckfort}}, \bibinfo {author}
  {\bibfnamefont {F.}~\bibnamefont {Hanke}}, \bibinfo {author} {\bibfnamefont
  {U.}~\bibnamefont {Pietrzyk}}, \bibinfo {author} {\bibfnamefont
  {K.}~\bibnamefont {Zilles}}, \bibinfo {author} {\bibfnamefont
  {K.}~\bibnamefont {Amunts}}, \ and\ \bibinfo {author} {\bibfnamefont
  {M.}~\bibnamefont {Axer}},\ }\href {\doibase
  10.1016/j.neuroimage.2015.02.020} {\bibfield  {journal} {\bibinfo  {journal}
  {Neuro{I}mage}\ }\textbf {\bibinfo {volume} {111}},\ \bibinfo {pages} {464}
  (\bibinfo {year} {2015})}\BibitemShut {NoStop}%
\bibitem [{\citenamefont {Menzel}\ \emph
  {et~al.}(2020{\natexlab{a}})\citenamefont {Menzel}, \citenamefont {Axer},
  \citenamefont {Raedt}, \citenamefont {Costantini}, \citenamefont {Silvestri},
  \citenamefont {Pavone}, \citenamefont {Amunts},\ and\ \citenamefont
  {Michielsen}}]{menzel2020}%
  \BibitemOpen
  \bibfield  {author} {\bibinfo {author} {\bibfnamefont {M.}~\bibnamefont
  {Menzel}}, \bibinfo {author} {\bibfnamefont {M.}~\bibnamefont {Axer}},
  \bibinfo {author} {\bibfnamefont {H.~D.}\ \bibnamefont {Raedt}}, \bibinfo
  {author} {\bibfnamefont {I.}~\bibnamefont {Costantini}}, \bibinfo {author}
  {\bibfnamefont {L.}~\bibnamefont {Silvestri}}, \bibinfo {author}
  {\bibfnamefont {F.~S.}\ \bibnamefont {Pavone}}, \bibinfo {author}
  {\bibfnamefont {K.}~\bibnamefont {Amunts}}, \ and\ \bibinfo {author}
  {\bibfnamefont {K.}~\bibnamefont {Michielsen}},\ }\href {\doibase
  10.1103/PhysRevX.10.021002} {\bibfield  {journal} {\bibinfo  {journal}
  {Physical Review X}\ }\textbf {\bibinfo {volume} {10}},\ \bibinfo {pages}
  {021002} (\bibinfo {year} {2020}{\natexlab{a}})}\BibitemShut {NoStop}%
\bibitem [{\citenamefont {Kumar}\ \emph {et~al.}(2014)\citenamefont {Kumar},
  \citenamefont {Petrik}, \citenamefont {Ramanandan}, \citenamefont {Gawhary},
  \citenamefont {Roy}, \citenamefont {Pereira}, \citenamefont {Coene},\ and\
  \citenamefont {Urbach}}]{kumar2014}%
  \BibitemOpen
  \bibfield  {author} {\bibinfo {author} {\bibfnamefont {N.}~\bibnamefont
  {Kumar}}, \bibinfo {author} {\bibfnamefont {P.}~\bibnamefont {Petrik}},
  \bibinfo {author} {\bibfnamefont {G.~K.~P.}\ \bibnamefont {Ramanandan}},
  \bibinfo {author} {\bibfnamefont {O.~E.}\ \bibnamefont {Gawhary}}, \bibinfo
  {author} {\bibfnamefont {S.}~\bibnamefont {Roy}}, \bibinfo {author}
  {\bibfnamefont {S.~F.}\ \bibnamefont {Pereira}}, \bibinfo {author}
  {\bibfnamefont {W.~M.~J.}\ \bibnamefont {Coene}}, \ and\ \bibinfo {author}
  {\bibfnamefont {H.~P.}\ \bibnamefont {Urbach}},\ }\href {\doibase
  10.1364/OE.22.024678} {\bibfield  {journal} {\bibinfo  {journal} {Optics
  Express}\ }\textbf {\bibinfo {volume} {22}},\ \bibinfo {pages} {24678}
  (\bibinfo {year} {2014})}\BibitemShut {NoStop}%
\bibitem [{\citenamefont {Gawhary}\ \emph {et~al.}(2011)\citenamefont
  {Gawhary}, \citenamefont {Kumar}, \citenamefont {Pereira}, \citenamefont
  {Coene},\ and\ \citenamefont {Urbach}}]{gawhary2011}%
  \BibitemOpen
  \bibfield  {author} {\bibinfo {author} {\bibfnamefont {O.~E.}\ \bibnamefont
  {Gawhary}}, \bibinfo {author} {\bibfnamefont {N.}~\bibnamefont {Kumar}},
  \bibinfo {author} {\bibfnamefont {S.~F.}\ \bibnamefont {Pereira}}, \bibinfo
  {author} {\bibfnamefont {W.~M.~J.}\ \bibnamefont {Coene}}, \ and\ \bibinfo
  {author} {\bibfnamefont {H.~P.}\ \bibnamefont {Urbach}},\ }\href {\doibase
  10.1007/s00340-011-4794-7} {\bibfield  {journal} {\bibinfo  {journal}
  {Applied Physics B}\ }\textbf {\bibinfo {volume} {105}},\ \bibinfo {pages}
  {775} (\bibinfo {year} {2011})}\BibitemShut {NoStop}%
\bibitem [{\citenamefont {Menzel}(2018)}]{MMenzel}%
  \BibitemOpen
  \bibfield  {author} {\bibinfo {author} {\bibfnamefont {M.}~\bibnamefont
  {Menzel}},\ }\href {\doibase 10.18154/RWTH-2018-230974} {\emph {\bibinfo
  {title} {{F}inite-difference time-domain simulations assisting to reconstruct
  the brain's nerve fiber architecture by 3{D} polarized light imaging}}},\
  \bibinfo {series} {Schriften des Forschungszentrums J\"ulich, Reihe
  Schl\"usseltechnologien}, Vol.\ \bibinfo {volume} {188}\ (\bibinfo
  {publisher} {Forschungszentrum J\"ulich GmbH},\ \bibinfo {address}
  {J\"ulich},\ \bibinfo {year} {2018})\BibitemShut {NoStop}%
\bibitem [{\citenamefont {Wilts}\ \emph {et~al.}(2014)\citenamefont {Wilts},
  \citenamefont {Michielsen}, \citenamefont {De~Raedt},\ and\ \citenamefont
  {Stavenga}}]{wilts2014}%
  \BibitemOpen
  \bibfield  {author} {\bibinfo {author} {\bibfnamefont {B.~D.}\ \bibnamefont
  {Wilts}}, \bibinfo {author} {\bibfnamefont {K.}~\bibnamefont {Michielsen}},
  \bibinfo {author} {\bibfnamefont {H.}~\bibnamefont {De~Raedt}}, \ and\
  \bibinfo {author} {\bibfnamefont {D.~G.}\ \bibnamefont {Stavenga}},\ }\href
  {\doibase 10.1073/pnas.1323611111} {\bibfield  {journal} {\bibinfo  {journal}
  {Proc. Natl. Acad. Sci.}\ }\textbf {\bibinfo {volume} {111}},\ \bibinfo
  {pages} {4363} (\bibinfo {year} {2014})}\BibitemShut {NoStop}%
\bibitem [{\citenamefont {Michielsen}\ \emph {et~al.}(2010)\citenamefont
  {Michielsen}, \citenamefont {de~Raedt},\ and\ \citenamefont
  {Stavenga}}]{michielsen2010}%
  \BibitemOpen
  \bibfield  {author} {\bibinfo {author} {\bibfnamefont {K.}~\bibnamefont
  {Michielsen}}, \bibinfo {author} {\bibfnamefont {H.}~\bibnamefont
  {de~Raedt}}, \ and\ \bibinfo {author} {\bibfnamefont {D.~G.}\ \bibnamefont
  {Stavenga}},\ }\href {\doibase 10.1098/rsif.2009.0352} {\bibfield  {journal}
  {\bibinfo  {journal} {J. Roy. Soc. Interface}\ }\textbf {\bibinfo {volume}
  {7}},\ \bibinfo {pages} {765} (\bibinfo {year} {2010})}\BibitemShut {NoStop}%
\bibitem [{\citenamefont {Taflove}\ and\ \citenamefont
  {Hagness}(2005)}]{taflove}%
  \BibitemOpen
  \bibfield  {author} {\bibinfo {author} {\bibfnamefont {A.}~\bibnamefont
  {Taflove}}\ and\ \bibinfo {author} {\bibfnamefont {S.~C.}\ \bibnamefont
  {Hagness}},\ }\href@noop {} {\emph {\bibinfo {title} {Computational
  Electrodynamics: The {F}inite-Difference {T}ime-Domain Method}}},\ \bibinfo
  {edition} {3rd}\ ed.\ (\bibinfo  {publisher} {Artech House},\ \bibinfo {year}
  {2005})\BibitemShut {NoStop}%
\bibitem [{\citenamefont {{Menzel}}\ \emph {et~al.}(2016)\citenamefont
  {{Menzel}}, \citenamefont {{Axer}}, \citenamefont {{De Raedt}},\ and\
  \citenamefont {{Michielsen}}}]{menzel2016}%
  \BibitemOpen
  \bibfield  {author} {\bibinfo {author} {\bibfnamefont {M.}~\bibnamefont
  {{Menzel}}}, \bibinfo {author} {\bibfnamefont {M.}~\bibnamefont {{Axer}}},
  \bibinfo {author} {\bibfnamefont {H.}~\bibnamefont {{De Raedt}}}, \ and\
  \bibinfo {author} {\bibfnamefont {K.}~\bibnamefont {{Michielsen}}},\ }in\
  \href {\doibase 10.1007/978-3-319-50862-7\_6} {\emph {\bibinfo {booktitle}
  {{Brain-Inspired Computing. BrainComp 2015. Lecture Notes in Computer
  Science}}}},\ Vol.\ \bibinfo {volume} {10087},\ \bibinfo {editor} {edited by\
  \bibinfo {editor} {\bibfnamefont {K.}~\bibnamefont {Amunts}}, \bibinfo
  {editor} {\bibfnamefont {L.}~\bibnamefont {Grandinetti}}, \bibinfo {editor}
  {\bibfnamefont {T.}~\bibnamefont {Lippert}}, \ and\ \bibinfo {editor}
  {\bibfnamefont {N.}~\bibnamefont {Petkov}}}\ (\bibinfo  {publisher} {Springer
  International Publishing},\ \bibinfo {address} {Cham},\ \bibinfo {year}
  {2016})\ Chap.~\bibinfo {chapter} {6}, pp.\ \bibinfo {pages}
  {73--85}\BibitemShut {NoStop}%
\bibitem [{\citenamefont {De~Raedt}(2005)}]{deRaedt}%
  \BibitemOpen
  \bibfield  {author} {\bibinfo {author} {\bibfnamefont {H.}~\bibnamefont
  {De~Raedt}},\ }in\ \href@noop {} {\emph {\bibinfo {booktitle} {Computational
  Electrodynamics: The Finite-Difference Time-Domain Method}}},\ \bibinfo
  {editor} {edited by\ \bibinfo {editor} {\bibfnamefont {A.}~\bibnamefont
  {Taflove}}\ and\ \bibinfo {editor} {\bibfnamefont {S.~C.}\ \bibnamefont
  {Hagness}}}\ (\bibinfo  {publisher} {Artech House},\ \bibinfo {address} {MA
  USA},\ \bibinfo {year} {2005})\ \bibinfo {edition} {3rd}\ ed.,\
  Chap.~\bibinfo {chapter} {18}\BibitemShut {NoStop}%
\bibitem [{\citenamefont {Savitsky}\ and\ \citenamefont
  {Golay}(1964)}]{savitsky1964}%
  \BibitemOpen
  \bibfield  {author} {\bibinfo {author} {\bibfnamefont {A.}~\bibnamefont
  {Savitsky}}\ and\ \bibinfo {author} {\bibfnamefont {M.}~\bibnamefont
  {Golay}},\ }\href {\doibase 10.1021/ac60214a047} {\bibfield  {journal}
  {\bibinfo  {journal} {Anal. Chem.}\ }\textbf {\bibinfo {volume} {36}},\
  \bibinfo {pages} {1627} (\bibinfo {year} {1964})}\BibitemShut {NoStop}%
\bibitem [{\citenamefont {Menzel}\ and\ \citenamefont
  {Pereira}(2020)}]{dataset}%
  \BibitemOpen
  \bibfield  {author} {\bibinfo {author} {\bibfnamefont {M.}~\bibnamefont
  {Menzel}}\ and\ \bibinfo {author} {\bibfnamefont {S.~F.}\ \bibnamefont
  {Pereira}},\ }\href {\doibase 10.17632/dp496jpd7h.2} {\bibfield  {journal}
  {\bibinfo  {journal} {Mendeley Data}\ }\textbf {\bibinfo {volume} {V4}},\
  \bibinfo {pages} {https://doi.org/10.17632/dp496jpd7h.4} (\bibinfo {year}
  {2020})}\BibitemShut {NoStop}%
\bibitem [{\citenamefont {Menzel}\ \emph
  {et~al.}(2020{\natexlab{b}})\citenamefont {Menzel}, \citenamefont {Huwer},
  \citenamefont {Schl\"{o}mer}, \citenamefont {Amunts},\ and\ \citenamefont
  {Axer}}]{menzel2020-OSA}%
  \BibitemOpen
  \bibfield  {author} {\bibinfo {author} {\bibfnamefont {M.}~\bibnamefont
  {Menzel}}, \bibinfo {author} {\bibfnamefont {M.}~\bibnamefont {Huwer}},
  \bibinfo {author} {\bibfnamefont {P.}~\bibnamefont {Schl\"{o}mer}}, \bibinfo
  {author} {\bibfnamefont {K.}~\bibnamefont {Amunts}}, \ and\ \bibinfo {author}
  {\bibfnamefont {M.}~\bibnamefont {Axer}},\ }in\ \href {\doibase
  10.1364/BRAIN.2020.BW2C.3} {\emph {\bibinfo {booktitle} {Biophotonics
  Congress: Biomedical Optics 2020 (Translational, Microscopy, OCT, OTS,
  BRAIN)}}}\ (\bibinfo  {publisher} {Optical Society of America},\ \bibinfo
  {year} {2020})\ p.\ \bibinfo {pages} {BW2C.3}\BibitemShut {NoStop}%
\bibitem [{\citenamefont {Jan}\ \emph {et~al.}(2015)\citenamefont {Jan},
  \citenamefont {Grimm}, \citenamefont {Tran}, \citenamefont {Lathrop},
  \citenamefont {Wollstein}, \citenamefont {Bilonick}, \citenamefont
  {Ishikawa}, \citenamefont {Kagemann}, \citenamefont {Schuman},\ and\
  \citenamefont {Sigal}}]{jan2015}%
  \BibitemOpen
  \bibfield  {author} {\bibinfo {author} {\bibfnamefont {N.-J.}\ \bibnamefont
  {Jan}}, \bibinfo {author} {\bibfnamefont {J.~L.}\ \bibnamefont {Grimm}},
  \bibinfo {author} {\bibfnamefont {H.}~\bibnamefont {Tran}}, \bibinfo {author}
  {\bibfnamefont {K.~L.}\ \bibnamefont {Lathrop}}, \bibinfo {author}
  {\bibfnamefont {G.}~\bibnamefont {Wollstein}}, \bibinfo {author}
  {\bibfnamefont {R.~A.}\ \bibnamefont {Bilonick}}, \bibinfo {author}
  {\bibfnamefont {H.}~\bibnamefont {Ishikawa}}, \bibinfo {author}
  {\bibfnamefont {L.}~\bibnamefont {Kagemann}}, \bibinfo {author}
  {\bibfnamefont {J.~S.}\ \bibnamefont {Schuman}}, \ and\ \bibinfo {author}
  {\bibfnamefont {I.~A.}\ \bibnamefont {Sigal}},\ }\href {\doibase
  10.1364/BOE.6.004705} {\bibfield  {journal} {\bibinfo  {journal} {Biomed.
  Opt. Express}\ }\textbf {\bibinfo {volume} {6}},\ \bibinfo {pages} {4705}
  (\bibinfo {year} {2015})}\BibitemShut {NoStop}%
\bibitem [{\citenamefont {Reuter}\ \emph {et~al.}(2019)\citenamefont {Reuter},
  \citenamefont {Matuschke}, \citenamefont {Menzel}, \citenamefont {Schubert},
  \citenamefont {Ginsburger}, \citenamefont {Poupon}, \citenamefont {Amunts},\
  and\ \citenamefont {Axer}}]{reuter2019}%
  \BibitemOpen
  \bibfield  {author} {\bibinfo {author} {\bibfnamefont {J.~A.}\ \bibnamefont
  {Reuter}}, \bibinfo {author} {\bibfnamefont {F.}~\bibnamefont {Matuschke}},
  \bibinfo {author} {\bibfnamefont {M.}~\bibnamefont {Menzel}}, \bibinfo
  {author} {\bibfnamefont {N.}~\bibnamefont {Schubert}}, \bibinfo {author}
  {\bibfnamefont {K.}~\bibnamefont {Ginsburger}}, \bibinfo {author}
  {\bibfnamefont {C.}~\bibnamefont {Poupon}}, \bibinfo {author} {\bibfnamefont
  {K.}~\bibnamefont {Amunts}}, \ and\ \bibinfo {author} {\bibfnamefont
  {M.}~\bibnamefont {Axer}},\ }\href {\doibase 10.1007/s11548-019-02053-6}
  {\bibfield  {journal} {\bibinfo  {journal} {International Journal of Computer
  Assisted Radiology and Surgery}\ }\textbf {\bibinfo {volume} {14}},\ \bibinfo
  {pages} {1881} (\bibinfo {year} {2019})}\BibitemShut {NoStop}%
\bibitem [{\citenamefont {Menzel}\ \emph {et~al.}(2019)\citenamefont {Menzel},
  \citenamefont {Axer}, \citenamefont {Amunts}, \citenamefont {Raedt},\ and\
  \citenamefont {Michielsen}}]{menzel2019}%
  \BibitemOpen
  \bibfield  {author} {\bibinfo {author} {\bibfnamefont {M.}~\bibnamefont
  {Menzel}}, \bibinfo {author} {\bibfnamefont {M.}~\bibnamefont {Axer}},
  \bibinfo {author} {\bibfnamefont {K.}~\bibnamefont {Amunts}}, \bibinfo
  {author} {\bibfnamefont {H.~D.}\ \bibnamefont {Raedt}}, \ and\ \bibinfo
  {author} {\bibfnamefont {K.}~\bibnamefont {Michielsen}},\ }\href {\doibase
  10.1038/s41598-019-38506-w} {\bibfield  {journal} {\bibinfo  {journal}
  {Scientific Reports}\ }\textbf {\bibinfo {volume} {9}},\ \bibinfo {pages}
  {1939} (\bibinfo {year} {2019})}\BibitemShut {NoStop}%
\bibitem [{\citenamefont {Menzel}\ \emph {et~al.}(2017)\citenamefont {Menzel},
  \citenamefont {Reckfort}, \citenamefont {Weigand}, \citenamefont {K\"{o}se},
  \citenamefont {Amunts},\ and\ \citenamefont {Axer}}]{menzel2017}%
  \BibitemOpen
  \bibfield  {author} {\bibinfo {author} {\bibfnamefont {M.}~\bibnamefont
  {Menzel}}, \bibinfo {author} {\bibfnamefont {J.}~\bibnamefont {Reckfort}},
  \bibinfo {author} {\bibfnamefont {D.}~\bibnamefont {Weigand}}, \bibinfo
  {author} {\bibfnamefont {H.}~\bibnamefont {K\"{o}se}}, \bibinfo {author}
  {\bibfnamefont {K.}~\bibnamefont {Amunts}}, \ and\ \bibinfo {author}
  {\bibfnamefont {M.}~\bibnamefont {Axer}},\ }\href {\doibase
  10.1364/BOE.8.003163} {\bibfield  {journal} {\bibinfo  {journal} {Biomed.
  Opt. Express}\ }\textbf {\bibinfo {volume} {8}},\ \bibinfo {pages} {3163}
  (\bibinfo {year} {2017})}\BibitemShut {NoStop}%
\end{thebibliography}%


\end{document}